\documentclass[letterpaper,twoside,titlepage,11pt]{article}
\usepackage{amssymb,amsmath,amsfonts} \usepackage{epsfig}

\newcommand{\be}{\begin{equation}} \newcommand{\ee}{\end{equation}}
\newcommand{\bea}{\begin{eqnarray}} \newcommand{\eea}{\end{eqnarray}}

\newcommand{\id}{\hbox{1\kern-.27em l}}
\newcommand{\sid}{\hbox{\scriptsize1\kern-.27em l}}

\newcommand{\we}{\kern-.1em\wedge\kern-.1em}
\newcommand{\scal}{\kern-.13em\cdot\kern-.13em}

\newcommand{\II}{I\kern-.09em I}

 \newcommand{\de}{\delta}
\newcommand{\ep}{\epsilon}

\newcommand{\Z}{\mathbb{Z}} 
\newcommand{\R}{\mathbb{R}}

\newcommand{\ads}{{\rm AdS}}

\newcommand{\nn}{\nonumber}
\newcommand{\spa}{\ \ ,\ \ \ \ }

\newcommand{\vecto}[2]{\left( \begin{array}{c} #1 \\ #2 \end{array}
\right) }

\newcommand{\ba}{\begin{eqnarray}}
\newcommand{\ea}{\end{eqnarray}}
\newcommand{\ca}{\mathcal}

\newcommand{\f}{\frac}
\newcommand{\s}{\sqrt}
\newcommand{\ti}{\tilde}
\newcommand{\ap}{\alpha}

\newcommand{\ddd}{\cdot\cdot\cdot}
\newcommand{\no}{\nonumber \\}

\newcommand{\di}{\partial}

\newcommand{\godel}{G\"{o}del }

\numberwithin{equation}{section}

\begin{document}

\begin{titlepage}

\rightline{\vbox{\small\hbox{HUTP-03/A004} }}
\vskip 2cm
\centerline{\LARGE\bf Supersymmetric \godel Universes in String Theory}
\vskip 1.2cm  \centerline{{\Large Troels Harmark and Tadashi Takayanagi}}
\vskip 0.7cm
\begin{center}
{\sl Jefferson Physical Laboratory\\ Harvard University\\
Cambridge, MA 02138, USA }
\vskip 0.7cm  
{\small \sffamily harmark@physics.harvard.edu \\

takayana@wigner.harvard.edu}
\end{center}
\vskip 1cm  \centerline{\bf \large Abstract} \vskip 0.5cm

\noindent 
Supersymmetric backgrounds in string and M-theory 
of the \godel Universe type are studied. 
We find several new \godel Universes that 
preserve up to 20 supersymmetries.
In particular, we obtain an interesting \godel Universe in M-theory with 18 
supersymmetries
which does not seem to be dual to a pp-wave.
We show that not only T-duality but also the 
type-IIA/M-theory 
S-duality can give supersymmetric \godel Universes from pp-waves.
We find solutions that can interpolate between \godel
Universes and pp-waves.
We also compute
the string spectrum on two type IIA \godel Universes. Furthermore, we  
obtain
the spectrum of D-branes on a \godel Universe and find
the supergravity solution for a D4-brane on a \godel Universe.

\end{titlepage}


\setcounter{page}{1}



\section{Introduction}

Many of the great breakthroughs in string theory in the past 
few years are tied to studying string theory on 
supersymmetric backgrounds with Ramond-Ramond fields.
Some of the most notable backgrounds include 
D-branes \cite{Polchinski:1995mt}, $\ads_5 \times S^5$ 
\cite{Maldacena:1998re,Aharony:1999ti} and, 
more recently, pp-waves with Ramond-Ramond fluxes 
\cite{Blau:2001ne,Metsaev:2001bj,Berenstein:2002jq}.

Recently, a new type of supersymmetric background
with Ramond-Ramond fluxes has been found. In 
\cite{Gauntlett:2002nw} an M-theory solution
of the \godel Universe type was found to preserve 20
supersymmetries.%
\footnote{Supersymmetric \godel Universes in compactified string theory
have been found in \cite{Horowitz:1995rf,Russo:1995cv}. These
solutions are not \godel Universes when uplifted to ten-dimensional
string theory.} 
This solution has Ramond-Ramond fluxes
when compactified to type IIA string theory.
The fact that it is of the \godel Universe
type means that it has closed time-like curves.
In \cite{Boyda:2002ba}, it was described that the
principle of Holography perhaps can remedy the problem of
closed time-like curves and protect the chronology in
\godel Universe backgrounds. 
Furthermore, in \cite{Boyda:2002ba}, it was shown that the 
\godel Universe background of \cite{Gauntlett:2002nw} is T-dual to a 
pp-wave.%
\footnote{This T-duality was also discussed in \cite{Herdeiro:2002ft}
but the connection to pp-waves was not considered.} 
This discovery gives a reason for the otherwise mysterious
existence of supersymmetric \godel Universes in that we can connect
them to already known solutions of string theory.

Originally the \godel universe \cite{Godel:1949ga} 
(see \cite{HawkingE} for an excellent review) is defined as a
pressure-free perfect fluid solution in General Relativity with 
negative cosmological constant. The four dimensional 
metric in polar coordinates is given by
\ba
\label{origodel}
ds^2=-dt^2+d\rho^2+\sinh^2 \rho \, (1-\sinh^2 \rho)d\phi^2
-2\s{2} \sinh^2 \rho \, dt d\phi + dz^2.\label{oldgodel}
\ea
We see that we have closed time-like curves for constant $\rho$ with
$\rho > \log(1+\s{2})$. Thus, the closed time-like curves cannot 
be arbitrarily small. The \godel Universe is also seen to be 
homogenous and to have a trivial topology.
On the other hand, for supersymmetric backgrounds of the \godel Universe
type in string theory 
and M-theory, a typical example of a metric is
\begin{equation}
\label{newgodel}
ds^2 = - dt^2 + d\rho^2 + \rho^2 ( 1 -\rho^2 ) d\phi^2 - 2\rho^2 dt d\phi,
\end{equation}
as a part of the ten or eleven-dimensional space-time.
Despite the different looking metric these supersymmetric
solutions still share the essential properties of the original
\godel Universe. 
Indeed this background is also homogenous, it has trivial topology,%
\footnote{Here we only mean the 3 dimensional part \eqref{newgodel}.
The remaining directions in string or M-theory
could be topologically non-trivial.} 
and we have closed time-like curves when $\rho > 1$.

In this paper we lay some of the ground work for a better understanding
of \godel Universes in string and M-theory. 
We first consider different ways of obtaining new \godel Universes,
and provide several new supersymmetric \godel Universe backgrounds.
We find that not only T-duality, but also the type-IIA/M-theory
S-duality can produce \godel Universes from pp-waves. 
We find the \godel Universes $G_{2n+1} \times \R^{10-2n}$ 
in M-theory with most supersymmetry. In particular, 
this includes a $G_{11}$ solution
with 18 supersymmetries, which is not related to the known background
by T-duality. 
We also find that there exist three inequivalent \godel Universes
with 20 supersymmetries. In this consideration we note a subtlety in 
the T-duality of \cite{Boyda:2002ba} on the maximally supersymmetric
type IIB pp-wave \cite{Blau:2001ne} 
which means that we can get two inequivalent 
type IIA \godel Universes, one with 20 supersymmetries
and one without supersymmetry.
Furthermore, we study two one-parameter families
of backgrounds which 
have the interesting feature that they interpolate between a \godel Universe
and a pp-wave.

We go on to study the string spectrum of the supersymmetric 
type IIA \godel Universe\cite{Boyda:2002ba} 
which is dual to the type IIB pp-wave
with maximal supersymmetry\cite{Blau:2001ne}.
To do this we use the fact that this pp-wave solution can be quantized
in the light-cone gauge\cite{Metsaev:2001bj}. We generalize the 
spectrum to include the compact direction on which we T-dualize
and thereby obtain the spectrum on the \godel Universe background.
We also find the spectrum for the type IIA \godel Universe
which is S-dual to the M-theory 
\godel Universe of \cite{Gauntlett:2002nw}.

Finally, we study D-branes on type IIA \godel Universes. We 
consider the boundary conditions and find the spectrum of D-branes.
We also find a new supergravity solution for a D4-brane on a \godel
Universe and uplift this to a new M5-brane solution on an M-theory
pp-wave.

\section{\godel Universes in String and M-theory}
\label{secsolutions}

In this section we consider the possible supersymmetric \godel Universes
in string and M-theory. We first consider the different ways of
obtaining supersymmetric \godel Universe solutions, either by 
T or S-dualizing
a supersymmetric pp-wave solution, or by directly looking for solutions
to the equations of motion and the Killing spinor equation. 
Then we go on to consider various new supersymmetric
\godel Universes and in particular we present the \godel Universes 
$G_{2n+1} \times \R^{10-2n}$ of M-theory for $n=1,2,3,4,5$
with most supersymmetry.

\subsection{\godel Universes from pp-waves by T-duality}
\label{godelbyT}

As shown in \cite{Boyda:2002ba} it is possible to find \godel Universe
backgrounds of type IIA/B string theory by T-dualizing 
type IIA/B pp-wave solutions. We explain here how this works in
general since we use this transformation repeatedly in the following.

Consider here a pp-wave metric of the form 
\begin{equation}
\label{genppII}
ds^2 = -2 dx^+ dx^-
- \beta^2 \sum_{k=1}^n a_k^2 \left[ (\tilde{x}^{2k-1})^2 +
(\tilde{x}^{2k})^2 \right] (dx^+)^2
+ \sum_{i=1}^{2n} (d\tilde{x}^i)^2 
+ \sum_{i=2n+1}^8 (dx^i)^2, 
\end{equation}
with $a_k \neq 0$. Our light-cone coordinates are defined by
$x^+ = t + y$ and $x^- = (t - y )/2$.

We assume here and in the following that we do not have NS-NS
flux turned on in this pp-wave background. In Section \ref{case2and4}
we give an
example where the NS-NS flux can hinder that we get a \godel Universe
after T-duality. 

We now do the coordinate transformation
\begin{equation}
\label{genct}
\tilde{x}^{2k-1} + i \tilde{x}^{2k}
= ( x^{2k-1} + i x^{2k} ) \exp \left( - i a_k \beta x^+ \right)
\ \ , \ \ k=1,...,n,
\end{equation}
In the new coordinate system we have 
\begin{equation}
\label{aftercoII}
ds^2 = - dt^2 + dy^2 + \sum_{i=1}^8 (dx^i)^2 
- 2\beta \sum_{i,j=1}^{2n} J_{ij} x^i dx^j (dt+dy),
\end{equation}
where we defined
\begin{equation}
\label{genJij}
J_{2k-1 , 2k} = - J_{2k , 2k-1} = a_k
\ \ , \ \ k=1,...,n.
\end{equation}
We see now that $y$ is an explicit space-like isometry of the 
metric \eqref{aftercoII}. We can therefore do a
T-duality along $y$. This gives the metric 
\begin{equation}
\label{metgodelII}
ds^2 = - \left( dt + \beta \sum_{i,j=1}^{2n} J_{ij} x^i dx^j \right)^2
+ \sum_{i=1}^8 (dx^i)^2 + dy^2,
\end{equation}
and the NS-NS 2-form potential $B_{(2)}$ and 3-form field strength
$H_{(3)}$
\begin{equation}
\label{NSNSfields}
B_{iy} = \beta \sum_{j=1}^{2n} J_{ij} x^j
\spa
H_{ijy} = - 2\beta J_{ij}
\ \ , \ \ i,j=1,...,2n.
\end{equation}
See Appendix \ref{appconv} for our T-duality conventions.
Clearly, the metric \eqref{metgodelII} is of the \godel Universe type,
i.e. it has closed time-like curves and a trivial topology,
and for $n=2$ and $n=4$ the metric is the same as for the \godel Universe
backgrounds found
in \cite{Gauntlett:2002nw,Boyda:2002ba}.

For any \godel Universe background that we consider in this paper
the time-space components of the metric $g_{0i}$, where $i$ runs
over all spatial directions (in Cartesian coordinates), 
can asymptotically be written as
$g_{0i} = \beta \sum_j J_{ij} x^j$. This gives the general definition
of the antisymmetric matrix $J_{ij}$. Also, we define
$n$ as being the rank of $J_{ij}$, i.e. $2n$ is the minimal
number of directions in which we can write the non-zero part of $J_{ij}$.

\subsection{\godel Universes from M-theory pp-waves by S-duality}
\label{godelbyS}

In this section we note that we can obtain \godel Universe backgrounds
by S-dualizing M-theory pp-waves. This is thus another way than
T-duality that pp-waves and \godel Universe backgrounds are dual.
In Section \ref{case2and4} we explain that the T- and S-dualities
can be mapped to each other when $n \leq 3$.

Consider the M-theory pp-wave metric 
\begin{equation}
\label{genppM}
ds^2 = -2 dx^+ dx^-
- \beta^2 \sum_{k=1}^n a_k^2 \left[ (\tilde{x}^{2k-1})^2 +
(\tilde{x}^{2k})^2 \right] (dx^+)^2
+ \sum_{i=1}^{2n} (d\tilde{x}^i)^2 
+ \sum_{i=2n+1}^9 (dx^i)^2, 
\end{equation}
with $a_k \neq 0$. For later convenience
we define here our light-cone coordinates by 
$x^+ = t + u$ and $x^- = (t - u)/2$.
Do now the coordinate transformation \eqref{genct}. This gives
the metric
\begin{equation}
\label{aftercoM}
ds^2 = - dt^2 + du^2 + \sum_{i=1}^9 (dx^i)^2 
- 2\beta \sum_{i,j=1}^{2n} J_{ij} x^i dx^j (dt+du),
\end{equation}
where $J_{ij}$ is as in \eqref{genJij}. Clearly we have
an explicit space-like isometry in the $u$ direction and
can therefore get the corresponding type IIA string theory
solution by compactifying this direction. 
We summarize the M/IIA S-duality rules in Appendix \ref{appconv}.
Using the relation 
\begin{equation}
ds_{\rm M}^2= ds_{\rm IIA}^2 + (du + A_\mu dx^\mu)^2,
\end{equation}
we get the ten-dimensional metric
\begin{equation}
\label{metgodelS}
ds^2 = - \left( dt + \beta \sum_{i,j=1}^{2n} J_{ij} x^i dx^j \right)^2
+ \sum_{i=1}^9 (dx^i)^2, 
\end{equation}
along with the gauge fields $A_{(1)}$ and the two-form Ramond-Ramond
(RR) field strength
$F_{(2)}$
\begin{equation}
\label{fluxgodelS}
A_i = \beta \sum_{j=1}^{2n} J_{ij} x^j
\spa
F_{ij} = - 2\beta J_{ij}
\ \ , \ \ i,j=1,...,2n.
\end{equation}
As we see from the metric we have thus obtained a 
\godel Universe background of type IIA string theory. In other words,
the strong coupling limit of the type IIA \godel Universe is given by
the pp-wave in M-theory.%
\footnote{Resolution of closed time-like curves by dimensional oxidation 
was also considered in \cite{Herdeiro:2000ap,Herdeiro:2002ft}.
In these cases the solutions were lifted from compactified string theory
to ten-dimensional string theory backgrounds.}
We use this S-duality below to find new supersymmetric \godel Universe
backgrounds of string and M-theory.

\subsection{Construction of M-theory \godel Universes}
\label{directgodel}

An alternative approach to get supersymmetric
\godel Universes instead of using T and S-dualities is to directly
find solutions of the equations of motions (EOMs) that
have supersymmetry. In this section we give the necessary tools
to do this for \godel Universes in M-theory.%
\footnote{See also \cite{Gauntlett:2002fz} for
general conditions on supersymmetric \godel Universe backgrounds.} 
These tools make a systematic study of supersymmetric \godel
Universes possible.
As we shall see in Section \ref{case5}, 
this approach can provide solutions
that we could not otherwise have obtained by use of dualities.

\subsubsection*{Ricci tensor for \godel Universes}

We consider a $G_{2n+1}$ \godel Universe with metric
\begin{equation}
\label{metgodeln}
ds^2 = - \left( dt + \sum_{i=1}^{n} c_i \rho_i^2 d\phi_i \right)^2
+ \sum_{i=1}^n \left( d\rho_i^2 + \rho_i^2 d\phi_i^2 \right),
\end{equation}
where $c_i$ are constants that characterize the metric.

First note that this metric is in polar coordinates contrary
to the metric of sections \ref{godelbyT} and \ref{godelbyS}.
To put the metrics of these sections in this form we should
perform the coordinate transformation
\begin{equation}
\label{polartrans}
x^{2k-1} + i x^{2k} = \rho_k e^{i \phi_k} \ \ , \ \ k=1,...,n.
\end{equation}
In the rest of this paper we shall always use this coordinate 
transformation when going between the Cartesian and polar coordinates.

Moreover, note that we are restricting ourselves to a special type
of \godel Universe metrics which is characterized by having
$g_{t\phi_i}$ proportional to $\rho_i^2$ since this seems to be the
right type of metrics to get supersymmetric \godel Universes.

The non-zero components of the Ricci tensor for the metric
\eqref{metgodeln} are
\begin{equation}
\label{riccin}
R^t_{\ t} = -2 \sum_{i=1}^n c_i^2,
\spa
R^{\rho_i}_{\ \rho_i} = R^{\phi_i}_{\ \phi_i} = 2 c_i^2,
\spa
R^t_{\ \phi_i} = - 2 c_i \rho_i^2 \left( c_i^2 + \sum_{j=1}^n c_j^2 \right).
\end{equation}
We note that curvature scalar is
\begin{equation}
\label{curvscal}
R = 2 \sum_{i=1}^n c_i^2.
\end{equation}
Using \eqref{riccin} one can write down the Einstein equations relating
the background fluxes to the geometry. We conduct this 
for M-theory \godel Universes
in the following.

\subsubsection*{Equations of motion for M-theory \godel Universes}

The bosonic part of 11 dimensional supergravity is
\begin{equation}
\label{11Dact}
I = \frac{1}{2\kappa^2} \int d^{11} x \sqrt{-g} \left(
R - \frac{1}{2 \cdot 4!} (F_{(4)})^2 \right)
- \frac{1}{6} \frac{1}{2\kappa^2} \int C_{(3)} \wedge F_{(4)} \wedge F_{(4)},
\end{equation}
with $F_{(4)} = dC_{(3)}$. This gives the Einstein equations
\begin{equation}
\label{einsteqs}
R^\mu_{\ \nu} = K^\mu_{\ \nu} \spa
K^\mu_{\ \nu} \equiv \frac{1}{2} \left( \frac{1}{3!} 
F^{\mu \sigma \xi \kappa} F_{\nu \sigma \xi \kappa}
- \frac{1}{3} \delta^\mu_\nu \frac{1}{4!} (F_{(4)})^2 \right).
\end{equation}
The equations of motion and Bianchi identity for the four-form field
strength are
\begin{equation}
\label{fseqs}
d * F_{(4)} = - \frac{1}{2} F_{(4)} \wedge F_{(4)}
\spa
d F_{(4)} = 0.
\end{equation}
We now consider the following ansatz for M-theory 
\godel Universes
\begin{equation}
\label{metgodel5}
ds^2 = - \left( dt + \sum_{i=1}^5 c_i \rho_i^2 d\phi_i \right)^2
+ \sum_{i=1}^5 \left( d\rho_i^2 + \rho_i^2 d\phi_i^2 \right),
\end{equation}
\begin{equation}
\label{fsansatz}
F_{\rho_i \phi_i \rho_j \phi_j} = \rho_i \rho_j a_{ij},
\end{equation}
with $i,j=1,...,5$, $c_i \geq 0$ and $a_{ij}$ a symmetric constant matrix
with $a_{ii}=0$. Thus we have ten independent constants in $a_{ij}$.
The $c_i$ are constants that 
we can take to be positive without loss of generality. 
To obtain a $G_{2n+1} \times R^{10-2n}$ \godel universe
one can put $c_{n+1} = \cdots = c_5 = 0$ and the other
$c_i$ non-zero.

We should also note that this is not the most general
ansatz one can have for supersymmetric \godel Universes in M-theory.
Indeed, in Section \ref{case2and4} we give two examples of supersymmetric
solutions that does not fit into the above ansatz. However, 
we believe that all the M-theory \godel Universes
with maximal supersymmetry, for a given $n$, should fit
into the above ansatz.

We first consider the Einstein equations. Define 
\begin{equation}
\label{defbi}
b_i = \sum_j a_{ij}^2.
\end{equation}
Then the non-zero components of $K^\mu_{\ \nu}$ are
\begin{equation}
K^t_{\ t} = - \frac{1}{12} \sum_i b_i
\spa
K^t_{\ \phi_i} = - \frac{1}{2} c_i \rho_i^2 b_i,
\spa
K^{\rho_i}_{\ \rho_i} = K^{\phi_i}_{\ \phi_i} = \frac{1}{2} b_i 
- \frac{1}{12} \sum_j b_j.
\end{equation}
By considering all the Einstein equations we can now see
that all EOMs are satisfied if
\begin{equation}
\label{getci}
c_i^2 = \frac{1}{4} b_i - \frac{1}{24} \sum_j b_j,
\end{equation}
or, equivalently, if
\begin{equation}
\label{getbi}
b_i = 4 c_i^2 + 4 \sum_j c_j^2. 
\end{equation}
Note that as consequence of equation \eqref{getci} the $c_i$'s 
are uniquely determined by giving the $a_{ij}$ matrix, since 
we imposed that $c_i \geq 0$.

Consider now the two remaining equations \eqref{fseqs}. 
The Bianchi identity is trivially fulfilled. The equations of
motion for the field strength can be written as
\begin{equation}
\label{fseom}
\partial_\sigma \left( \sqrt{-g} F^{\sigma \mu \nu \xi} \right)
= \frac{1}{2} \frac{1}{(4!)^2} 
\epsilon^{\mu \nu \xi \kappa_1 \cdots \kappa_8}\
F_{\kappa_1 \cdots \kappa_4} F_{\kappa_5 \cdots \kappa_8}, 
\end{equation}
where $\epsilon^{01 \cdots \sharp} =1$ (we denote the eleventh
coordinate by $x^\sharp$). Using that
$\sqrt{-g} = \rho_1 \cdots \rho_5$ we get that \eqref{fseom}
is satisfied if%
\footnote{The factor of $1/4$ on the RHS is present because this means
that e.g. $ \frac{1}{4} \sum_{j,k,l,m} \eta^{1jklm} a_{jk} a_{lm}
= a_{23} a_{45} + a_{24} a_{35} + a_{25} a_{34}$.}
\begin{equation}
\label{eomfromfs}
- 2 \sum_j a_{ij} c_j = \frac{1}{4} \sum_{j,k,l,m} 
\eta^{ijklm} a_{jk} a_{lm},
\end{equation}
where
\begin{equation}
\label{etasymbol}
\eta^{ijklm} = \left\{ \begin{array}{rl} 1 & \mbox{ if } i,j,k,l,m 
\mbox{ are all
different} \\ 0 & \mbox{ otherwise} \end{array} \right..
\end{equation}
Thus, for a given collection of $c_i$ and $a_{ij}$ 
the EOMs of M-theory reduce to the equations \eqref{getci} 
and \eqref{eomfromfs}. 
Since \eqref{getci} determines
the $c_i$ uniquely, we can see \eqref{eomfromfs} as a condition
that a given $a_{ij}$ matrix should fulfil in order to correspond
to a solution.

Note that if we start with the $c_i$'s given, we can try to solve
for $a_{ij}$. Since we have ten independent components of $a_{ij}$
and ten independent equations \eqref{getbi} and \eqref{eomfromfs} 
this looks possible. This is not the case, however, there can be 
many different matrices $a_{ij}$ giving the same $c_i$'s.

For use in the rest of the paper, we introduce here 
a shorthand expression for a given \godel Universe solution 
\begin{equation}
(c_1,c_2,c_3,c_4,c_5) \spa 
(a_{12},a_{13},a_{14},a_{15},a_{23},a_{24},a_{25},a_{34},a_{35},a_{45}),
\end{equation}
that specifies uniquely an M-theory \godel Universe background.

\subsubsection*{Supersymmetry conditions for M-theory \godel Universes}

To count the number of supersymmetries of a classical 
solution in M-theory, we need
to count the number of independent spinors $\eta$ that obey the 
Killing spinor equation
\begin{equation}
\label{killspin}
D_\mu \eta + \frac{1}{288} \left( \Gamma_\mu^{\ \nu \sigma \kappa \xi}
- 8 \delta_\mu^\nu \Gamma^{\sigma\kappa\xi} \right) 
\eta F_{\nu\sigma\kappa\xi} = 0,
\end{equation}
with $D_\mu \eta = \partial_\mu \eta + \frac{1}{4} \omega_{\mu ab} 
\Gamma^{ab} \eta$. The Gamma-matrices $\Gamma^\mu$ are for the 
curved space, while $\Gamma^{a}$ are for the tangent space. 

For our ansatz \eqref{metgodel5}-\eqref{fsansatz} it is convinient
to classify constant spinors according to their eigenvalues under the
matrices
\begin{equation}
\label{theJs}
J_1 = \Gamma^{012} \spa
J_2 = \Gamma^{034} \spa
J_3 = \Gamma^{056} \spa
J_4 = \Gamma^{078} \spa
J_5 = \Gamma^{09\sharp}.
\end{equation}
Here the Gamma matrices are for the tangent space. 
The eigenvalues of $J_i$ are $\pm 1$.
Then since we impose $(\Gamma_{\sharp})^2 = \Gamma_{01 \cdots \sharp} = 1$
we have the constraint
\begin{equation}
\label{Jcond}
\prod_{i=1}^5 J_i = -1,
\end{equation}
Thus, we can classify constant spinors
according to their eigenvalues of $(J_1,J_2,J_3,J_4)$.
Note that any eigenvalue of $(J_1,J_2,J_3,J_4)$ has degeneracy two.

Our ansatz for a Killing spinor obeying
\eqref{killspin} is
\begin{equation}
\eta^{(s_1,s_2,s_3,s_4)} 
= \left[ 1 + \sum_{k=1}^5 u_k^{(s_1,s_2,s_3,s_4)} 
\Gamma^0 \left( \Gamma^{2k} x^{2k-1} - \Gamma^{2k-1} x^{2k} \right) \right]
\eta_0^{(s_1,s_2,s_3,s_4)},
\end{equation}
where $\eta_0^{(s_1,s_2,s_3,s_4)}$ is a constant spinor with
eigenvalues $(s_1,s_2,s_3,s_4)$ of $(J_1,J_2,J_3,J_4)$.
The Gamma matrices in this expression are flat.
The Killing spinor equation \eqref{killspin} then becomes
\begin{equation}
\label{timecomp1}
\left( 6 \sum_{i=1} c_i J_i - \sum_{i<j} a_{ij} J_i J_j \right) 
\eta^{(s_1,s_2,s_3,s_4)} = 0,
\end{equation}
for the time-component, and 
\begin{equation}
\label{spacecomp1}
\partial_{2k-1} \eta^{(s_1,s_2,s_3,s_4)}
= \Gamma^0 \Gamma^{2k} L \eta^{(s_1,s_2,s_3,s_4)}
\spa 
\partial_{2k} \eta^{(s_1,s_2,s_3,s_4)}
= - \Gamma^0 \Gamma^{2k-1} L \eta^{(s_1,s_2,s_3,s_4)},
\end{equation}
\begin{equation}
L \equiv - \frac{1}{12} J_k \sum_{i<j} a_{ij} J_i J_j 
+ \frac{1}{4} \sum_{i=1}^5 a_{ki} J_i - \frac{1}{2} c_k,
\end{equation}
for the spatial components. This reduces now to the equations
\begin{equation}
\label{susycount}
6 \sum_{i=1} c_i s_i = \sum_{i<j} a_{ij} s_i s_j  
\spa
L \left( \eta^{(s_1,s_2,s_3,s_4)} - \eta_0^{(s_1,s_2,s_3,s_4)} \right) = 0,
\end{equation}
provided we take
\begin{equation}
u_k^{(s_1,s_2,s_3,s_4)} = - \frac{1}{12} s_k \sum_{i<j} a_{ij} s_i s_j 
+ \frac{1}{4} \sum_{i=1}^5 a_{ki} s_i - \frac{1}{2} c_k.
\end{equation}
In conclusion,
the number of supersymmetries for a given
choice of $c_i$ and $a_{ij}$ is given by two times the 
number of solutions to \eqref{susycount} when trying the 16 possible
choices of $(s_1,s_2,s_3,s_4)$.

\subsubsection*{Symmetries of the ansatz}

We can also consider the possible symmetries of the ansatz
\eqref{metgodel5}-\eqref{fsansatz} for a given \godel Universe solution. 
I.e. for given $c_i$ and $a_{ij}$, what general transformations
can give $\tilde{c}_i$ and
$\tilde{a}_{ij}$ that also solves the EOMs \eqref{getci}
and \eqref{eomfromfs}, and have the same amount of supersymmetry 
(which is computed from \eqref{susycount}). We have 
three basic transformations
that map solutions to solutions keeping the same amount of supersymmetry.

The first one is that we can rescale the $c_i$ and $a_{ij}$, i.e.
the transformation is 
$\tilde{a}_{ij} = \gamma a_{ij}$ and $\tilde{c}_i = \gamma c_i$ 
with $\gamma \neq 0$.

The second one is that we interchange two of the five different two-planes.
If we for example exchange the 12-plane and 34-plane we should make
the transformation 
$\tilde{a}_{12} = a_{12}$, 
$\tilde{a}_{i1} = a_{i2}$, 
$\tilde{a}_{i2} = a_{i1}$ 
and 
$\tilde{c}_1 = c_2$, 
$\tilde{c}_2 = c_1$, 
$\tilde{c}_i=c_i$ with $i,j = 3,4,5$.

The third transformation is that we can change parity in one of the
planes. If we for example change parity in the $9\sharp$-plane we 
should make the transformation
$\tilde{a}_{ij} = - a_{ij}$, $\tilde{a}_{i5} = a_{i5}$
and
$\tilde{c}_i = c_i$,
$\tilde{c}_5 = - c_5$ 
with $i,j=1,2,3,4$.

\subsection{Solutions with $n=2$ and $n=4$}
\label{case2and4}

\subsubsection*{$n=2$ \godel Universes}

In \cite{Gauntlett:2002nw} an
$n=2$ \godel Universe background of M-theory with 20 supersymmetries
was found. In \cite{Boyda:2002ba} this was shown to be T-dual
to a pp-wave with 24 supersymmetries. We review this briefly and
put the T-duality into a broader picture involving also the
S-duality transformation of Section \ref{godelbyS}.

In \cite{Cvetic:2002hi} the Penrose limit of $\ads_3 \times S^3 \times \R^4$
from intersecting D3-branes was found to be the type IIB pp-wave background
with the metric
\begin{equation}
\label{metppintD3}
ds^2 = - 2 dx^+ dx^- - \beta^2 \sum_{i=1}^4 (\tilde{x}^i)^2 (dx^+)^2
+ \sum_{i=1}^4 (d\tilde{x}^i)^2
+ \sum_{i=5}^8 (dx^i)^2,
\end{equation}
and the RR fields
\begin{equation}
\label{fluxppintD3}
F^{(5)} = - 2\beta dx^+ ( d\tilde{x}^1 d\tilde{x}^2 
+ d\tilde{x}^3 d\tilde{x}^4) ( dx^5 dx^6 + dx^7 dx^8 ).
\end{equation}
We see the metric corresponds to $a_1=a_2=1$ in Section \ref{godelbyT}.
After the T-duality transformation of Section \ref{godelbyT} we get
the \godel Universe background of type IIA
\begin{equation}
\label{G5IIA}
ds^2 = - \left( dt + \beta \sum_{i,j=1}^{4} J_{ij} x^i dx^j \right)^2
+ \sum_{i=1}^8 (dx^i)^2 + dy^2,
\end{equation}
\begin{equation}
\label{fluxG5IIA}
H_{12y} = H_{34y} = - 2\beta 
\spa
F_{1256} = F_{1278} = F_{3456} = F_{3478} = - 2\beta,
\end{equation}
with $J_{12} = J_{34} = 1$.
Lifting to M-theory we get the $G_5 \times \R^6$ \godel Universe
background
\begin{equation}
\label{metn2M}
ds^2 = - \left( dt + \beta \sum_{i,j=1}^{4} J_{ij} x^i dx^j \right)^2
+ \sum_{i=1}^{\sharp} (dx^i)^2, 
\end{equation}
\begin{equation}
\label{fluxn2M}
F_{1256} = F_{1278} = F_{3456} = F_{3478} 
= F_{129\sharp} = F_{349\sharp}
= - 2\beta,
\end{equation}
where $x^9=y$. 
This M-theory \godel Universe background has 20 supersymmetries
\cite{Gauntlett:2002nw}.%
\footnote{Notice that the number of supersymmetries 
can be changed under T-duality as we examine in Appendix \ref{appsusy}.}
In our notation of Section \ref{directgodel}
we can write this solution as
\begin{equation}
(1,1,0,0,0) \spa (0,-2,-2,-2,-2,-2,-2,0,0,0).
\end{equation}

If we T-dualize the pp-wave background 
\eqref{metppintD3}-\eqref{fluxppintD3} in the $x^7$ and $x^8$ directions
we have the same metric but the RR fields are given by
\begin{equation}
\label{fluxppD1D5}
F_{(3)} = 2\beta dx^+ ( d\tilde{x}^1 d\tilde{x}^2 
+ d\tilde{x}^3 d\tilde{x}^4 ).
\end{equation}
This type IIB pp-wave with 24 supersymmetries can be obtained from 
a Penrose limit of 
$\ads_3 \times S^3 \times \R^4$ from the D1-D5 system 
\cite{Berenstein:2002jq,Russo:2002rq,Cvetic:2002si,Gomis:2002qi,Bena:2002kq}. 
If we do the T-duality of Section \ref{godelbyT}
we get the \godel Universe 
metric \eqref{G5IIA} and the NS-NS and RR fields
\begin{equation}
\label{fluxG5IIA2}
H_{12y} = H_{34y} = - 2\beta 
\spa
F_{12} = F_{34} = - 2\beta 
\spa
F_{012y} = F_{034y} = 2\beta,
\end{equation}
This \godel Universe background of type IIA has also 20 supersymmetries.
Indeed this \godel Universe background is related by T-dualities in the
$x^7$ and $x^8$ directions to the background 
\eqref{G5IIA}-\eqref{fluxG5IIA}.

We now uplift the type IIA \godel Universe \eqref{G5IIA},
\eqref{fluxG5IIA2} to M-theory. We call $y=x^9$ and the eleventh
direction $u$. We then get
\begin{equation}
\label{metmpp24}
ds^2 = - dt^2 + du^2 + \sum_{i=1}^9 (dx^i)^2 
- 2\beta \sum_{i,j=1}^{4} J_{ij} x^i dx^j (dt+du),
\end{equation}
\begin{equation}
\label{fluxmpp24}
F_{0129} = F_{u129} = F_{0349} = F_{u349} = 2\beta.
\end{equation}
This is an M-theory pp-wave solution with 24 supersymmetries, obtained from
a Penrose limit in \cite{Cvetic:2002si}. 
We see that we have connected the type IIA \godel Universe \eqref{G5IIA},
\eqref{fluxG5IIA2} to the M-theory pp-wave
\eqref{metmpp24}-\eqref{fluxmpp24} by the S-duality transformation
described in Section \ref{godelbyS}.

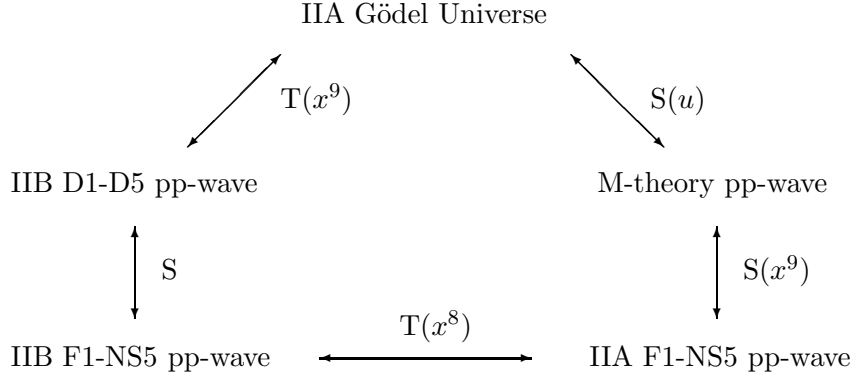
\begin{figure}[h]
\begin{picture}(390,160)(-15,10)
\put(85,35){\vector(0,1){35}}
\put(85,70){\vector(0,-1){35}}
\put(305,35){\vector(0,1){35}}
\put(305,70){\vector(0,-1){35}}
\put(155,20){\vector(1,0){80}}
\put(235,20){\vector(-1,0){80}}
\put(105,100){\vector(1,1){35}}
\put(140,135){\vector(-1,-1){35}}
\put(285,100){\vector(-1,1){35}}
\put(250,135){\vector(1,-1){35}}
\put(148,147){\shortstack[l]{IIA \godel Universe}}
\put(38,83){\shortstack[l]{IIB D1-D5 pp-wave}}
\put(38,17){\shortstack[l]{IIB F1-NS5 pp-wave}}
\put(257,17){\shortstack[l]{IIA F1-NS5 pp-wave}}
\put(260,83){\shortstack[l]{M-theory pp-wave}}
\put(140,115){\shortstack[l]{$\mbox{T}(x^9)$}}
\put(95,50){\shortstack[l]{$\mbox{S}$}}
\put(280,115){\shortstack[l]{$\mbox{S}(u)$}}
\put(315,50){\shortstack[l]{$\mbox{S}(x^9)$}}
\put(185,29){\shortstack[l]{$\mbox{T}(x^8)$}}
\end{picture}
\caption{Diagram that shows connection between T and S-duality
transformations relating \godel Universes and pp-waves. \label{figTS} }
\end{figure}

We can now demonstrate that the T-duality transformation of
Section \ref{godelbyT} and the S-duality transformation of
Section \ref{godelbyS} in fact are equivalent when
we have at least two flat directions, i.e. when $n \leq 3$ for
the \godel Universes. 
Consider the M-theory pp-wave \eqref{metmpp24}-\eqref{fluxmpp24}.
Compactify now this on $T^3$ in the $x^8$, $y=x^9$ and the $u$ directions.
We have then illustrated in Figure \ref{figTS} a duality chain that
connects the T-duality of Section \ref{godelbyT} and the S-duality of
Section \ref{godelbyS}. 
Starting from the M-theory pp-wave \eqref{metmpp24}-\eqref{fluxmpp24}
we can S-dualize along $y=x^9$ to obtain the type IIA pp-wave
which comes from a Penrose limit of 
F1-NS5 with 24 supersymmetries \cite{Russo:2002rq,Hikida:2002in}. 
This we can T-dualize 
along $x^8$ to obtain the F1-NS5 pp-wave in type IIB. Using
type IIB S-duality we arrive at the D1-D5 pp-wave 
\eqref{metppintD3}, \eqref{fluxppD1D5}. Then we do the T-duality
of Section \ref{godelbyT} along $y=x^9$ to obtain 
the type IIA \godel Universe \eqref{G5IIA}, \eqref{fluxG5IIA2}.
Finally, we end up with the M-theory pp-wave 
\eqref{metmpp24}-\eqref{fluxmpp24} by uplift to M-theory.
In conclusion we have shown that for $n \leq 3$
the T and S-duality that
relate pp-waves to \godel Universes are equivalent for M-theory on $T^3$.

\subsubsection*{$n=2$ \godel Universe/pp-wave mixture}

It is interesting to consider the mixed pp-wave background
of type IIB with 
metric \eqref{metppintD3} and NS-NS and RR three form fluxes
\cite{Berenstein:2002jq}
\begin{equation}
\label{fluxppmix}
F_{(3)} = 2\beta \cos \gamma \, dx^+ ( d\tilde{x}^1 d\tilde{x}^2 
+ d\tilde{x}^3 d\tilde{x}^4 )
\spa
H_{(3)} = - 2\beta \sin \gamma \, dx^+ ( d\tilde{x}^1 d\tilde{x}^2 
+ d\tilde{x}^3 d\tilde{x}^4 ).
\end{equation}
This background has 24 supersymmetries. 
We now T-dualize this according to Section \ref{godelbyT}
with $a_1=a_2=1$. We subsequently do two trivial T-dualities along
$x^8$ and $x^7$, and then make a trivial uplift to M-theory.
The resulting M-theory background is
\begin{equation}
\label{metn2mix}
ds^2 = - \left( dt + \beta \sum_{i,j=1}^{4} J_{ij} x^i dx^j \right)^2
+ \left( dy - \beta \sin \gamma \sum_{i,j=1}^{4} J_{ij} x^i dx^j \right)^2 
+ \sum_{i=1}^9 (dx^i)^2, 
\end{equation}
\begin{eqnarray}
\label{fluxn2mix}
& F_{0129} = F_{0349} = - 2\beta \sin \gamma
\spa
F_{y129} = F_{y349} = - 2\beta, & \nn \\
& F_{1256} = F_{1278} = F_{3456} = F_{3478} = - 2\beta \cos \gamma, &
\end{eqnarray}
where $J_{12}=J_{34} = 1$ and $x^9$ is the eleventh direction.

This solution is interesting for several reasons. If we consider
the special case $\gamma=0$ it reduces to the
M-theory $n=2$ \godel Universe \eqref{metn2M}-\eqref{fluxn2M}.
On the other hand, if we consider the case $\gamma=\pi/2$
it becomes an M-theory pp-wave with 24 supersymmetries \cite{Cvetic:2002si}.
The solution \eqref{metn2mix}-\eqref{fluxn2mix} thus interpolates
between being a \godel Universe and a pp-wave.

Moreover, we see that for $\gamma = \pi/2$ we have an example of a 
supersymmetric pp-wave solution with NS-NS flux that does not
become a \godel Universe background after the T-duality of
Section \ref{godelbyT}.

In general the solution \eqref{metn2mix}-\eqref{fluxn2mix} has 
16 supersymmetries while, as we just saw, it has 20 supersymmetries 
when $\gamma = 0$ and 24 supersymmetries when $\gamma = \pi/2$.
We give a proof of the 16 supersymmetries of the solution
in Appendix \ref{appsusy}.

Note that the solution \eqref{metn2mix}-\eqref{fluxn2mix}
does not fit into the ansatz \eqref{metgodel5}-\eqref{fsansatz}
since the $g_{yi}$ and the $F_{0ijk}$ components are non-zero.
This is of course in accordance with the fact that the solution
\eqref{metn2mix}-\eqref{fluxn2mix} can describe both a pp-wave and
a \godel Universe.

To understand better the interpolation between \godel Universe and pp-wave
we write the metric \eqref{metn2mix} in polar coordinates
\begin{eqnarray}
ds^2 &=& - dt^2 + dy^2 + \sum_{i=5}^9 (dx^i)^2
+ \sum_{i=1}^2 \Big[ d\rho_i^2 
+ \rho_i^2 \left( 1 - \beta^2 \cos^2 \gamma \, \rho_i^2 \right)
d\phi_i^2 \Big] \nn \\
&&
- 2 \beta^2 \cos^2 \gamma \, \rho_1^2 \rho_2^2 d\phi_1 d\phi_2
- 2 \beta \left( dt + \sin \gamma \, dy \right) 
\left( \rho_1^2 d\phi_1 + \rho_2^2 d\phi_2 \right).
\end{eqnarray}
If we consider for example the 12-plane, we see that for 
the curve $\rho_1 = \mbox{constant}$ it is a closed space-like curve
for $\rho_1 < 1 / ( \beta \cos \gamma )$ and a closed
time-like curve for $\rho_1 > 1 / ( \beta \cos \gamma )$.
If we now consider the limit $\cos \gamma \rightarrow 0$ we see
that the necessary radius to make a closed time-like curves goes
to infinity. So if $\cos \gamma$ is very small but non-zero the geometry is
almost that of a pp-wave and the radii of the closed time-like curves
are so large that they effectively can be ignored. 
It would
be interesting to study how the holographic sheets 
behave in this limit.

\subsubsection*{$n=4$ \godel Universes}

In \cite{Boyda:2002ba} it was pointed out that 
the maximally supersymmetric pp-wave of
type IIB \cite{Blau:2001ne} 
can be T-dualized into an $n=4$ \godel Universe solution.
In the following we examine this T-duality carefully 
and study its supersymmetry. We find that the T-duality can
give two inequivalent $n=4$ \godel Universes. We then
go on to discuss a more general type of pp-wave solution and
the consequences this solution have for the relation between
the T and S-dualities of Sections \ref{godelbyT} and \ref{godelbyS}.
In particular we find a new \godel Universe with 20 supersymmetries.

Consider the maximally symmetric pp-wave of type IIB \cite{Blau:2001ne}
with metric
\begin{equation}
\label{metmaxpp}
ds^2 = - 2 dx^+ dx^- - \beta^2 \sum_{i=1}^8 (\tilde{x}^i)^2 
(dx^+)^2 + \sum_{i=1}^8 (d\tilde{x}^i)^2, 
\end{equation}
and RR flux
\begin{equation}
\label{fluxmaxpp}
F_{(5)} = 4 \beta dx^+ \left(d\tilde{x}^1 d\tilde{x}^2 d\tilde{x}^3 
d\tilde{x}^4 + d\tilde{x}^5 d\tilde{x}^6 d\tilde{x}^7 d\tilde{x}^8 \right).
\end{equation}
We see that the metric corresponds to 
$ a_1^2 = a_2^2 = a_3^2 = a_4^2 = 1$ in Section \ref{godelbyT}. 
However, we shall see in the following that the signs of the $a_i$'s
become important. Without loss of generality we therefore choose
$a_1=a_2=a_3=1$ and $a_4 = s$, where $s=\pm 1$.

We consider now the T-duality of Section \ref{godelbyT}. 
We get the metric
\begin{equation}
\label{metn4iia1}
ds^2 = - \left( dt + \beta \sum_{i,j=1}^{8} J_{ij} x^i dx^j \right)^2
+ \sum_{i=1}^8 (dx^i)^2 + dy^2,
\end{equation}
with $J_{12} = J_{34} = J_{56} = 1$
and $J_{78} = s$.
The NS-NS and RR fluxes are
\begin{equation}
\label{fluxn4iia1}
H_{12y} = H_{34y} = H_{56y} = - 2\beta
\spa
H_{78y} = - 2 s \beta 
\spa
F_{1234} = F_{5678} = 4 \beta.
\end{equation}
As we shall see below, this defines two inequivalent 
$n=4$ type IIA \godel Universe backgrounds, corresponding to $s=\pm 1$.

We can now uplift these solutions to M-theory. This gives the $n=4$
M-theory \godel Universes 
\begin{equation}
\label{metn4M1}
ds^2 = - \left( dt + \beta \sum_{i,j=1}^{8} J_{ij} x^i dx^j \right)^2
+ \sum_{i=1}^\sharp (dx^i)^2, 
\end{equation}
\begin{equation}
\label{fluxn4M1}
F_{129\sharp} = F_{349\sharp} = F_{569\sharp} = - 2\beta
\spa
F_{789\sharp} = - 2 s \beta 
\spa
F_{1234} = F_{5678} = 4 \beta,
\end{equation}
with $y=x^9$ as usual. In our short hand notation of 
Section \ref{directgodel} we can write the solutions as
\begin{equation}
(1,1,1,s,0) \spa (4,0,0,-2,0,0,-2,4,-2,-2s).
\end{equation}
Using the methods in Section \ref{directgodel}
it is now easy to compute that the number of supersymmetries is
zero if $s=1$ and 20 if $s=-1$.
For the detail of Killing spinors see (\ref{ks420}) in the Appendix B.
This is obviously also true for the type IIA \godel Universes
\eqref{metn4iia1}-\eqref{fluxn4iia1}. We also computed the 20
supersymmetries of the $s=-1$ type IIA solution in an alternative way in
Appendix \ref{appsusy}.
We denote the $s=-1$ M-theory \godel Universe as $G_9 \times R^2$.

If we want to write down the $s=-1$ solution with $c_4=1$ we note that
using the parity changing 
transformation of Section \ref{directgodel} we get the
equivalent solution
\begin{equation}
(1,1,1,1,0) \spa (-4,0,0,2,0,0,2,4,2,2),
\end{equation}
with 20 supersymmetries.

We now go on to study a one-parameter 
type IIB pp-wave solution with 28 supersymmetries and its dual type IIA
and M-theory \godel Universe. As we shall see, this illustrates
that the T and S dualities in Section
\ref{godelbyT} and \ref{godelbyS} are not equivalent for $n = 4$. 
It also has other interesting features.

The type IIB pp-wave solution with 28 supersymmetries \cite{Bena:2002kq}
(see also \cite{Michelson:2002ps}) has metric
\begin{equation}
\label{metpp28}
ds^2 = - 2dx^+ dx^- - \beta^2 \sum_{i=1}^8 (\tilde{x}^i)^2 
(dx^+)^2 + \sum_{i=1}^8 (d\tilde{x}^i)^2, 
\end{equation}
and RR fluxes
\begin{eqnarray}
\label{fluxpp28}
& F_{(3)} = 2\beta \sin \gamma dx^+ \left( d\tilde{x}^1 d\tilde{x}^2
+ d\tilde{x}^3 d\tilde{x}^4 + d\tilde{x}^5 d\tilde{x}^6
- d\tilde{x}^7 d\tilde{x}^8 \right), & \nn \\
& F_{(5)} = 4 \beta \cos \gamma 
dx^+ \left(d\tilde{x}^1 d\tilde{x}^2 d\tilde{x}^3 
d\tilde{x}^4 + d\tilde{x}^5 d\tilde{x}^6 d\tilde{x}^7 d\tilde{x}^8 \right), &
\end{eqnarray}
$\gamma$ is the parameter of the solution.
When $\gamma=0$ we regain the maximally supersymmetric pp-wave of type IIB
\cite{Blau:2001ne}. When $\gamma=\pi/2$ 
we get a type IIB pp-wave with 28 
supersymmetries found in \cite{Bena:2002kq,Michelson:2002ps}.

We now take the T-duality in Section \ref{godelbyT}. 
{}From the $\gamma=0$ case above we know that to get a supersymmetric
solution we should take $a_1 = a_2 = a_3 = - a_4 = 1$.
The T-duality then gives the type IIA \godel Universe background
\begin{equation}
\label{metn4iia2}
ds^2 = - \left( dt + \beta \sum_{i,j=1}^{8} J_{ij} x^i dx^j \right)^2
+ \sum_{i=1}^8 (dx^i)^2 + dy^2,
\end{equation}
with 
$J_{12} = J_{34} = J_{56} = 1$
and $J_{78} = - 1$.
The NS-NS and RR fluxes are
\begin{eqnarray}
\label{fluxn4iia2}
& F_{12} = F_{34} = F_{56} = - F_{78} = - 2 \beta \sin \gamma \spa
F_{012y} = F_{034y} = F_{056y} = - F_{078y} = 2 \beta \sin \gamma, & \nn \\ 
& H_{12y} = H_{34y} = H_{56y} = - H_{78y} = - 2 \beta
\spa
F_{1234} = F_{5678} = 4 \beta \cos \gamma. 
 & 
\end{eqnarray}
This type IIA \godel Universe solution has generically 12 supersymmetries,
(see Appendix \ref{appsusy} for a proof of this).
In the two special cases $\gamma = 0 $ and $\gamma = \pi/2$ it has
instead 20 supersymmetries (see Appendix \ref{appsusy} for the
$\gamma = \pi/2$ case). Therefore we see that we have two different
$n=4$ \godel Universes of type IIA with
20 supersymmetries which are not related by dualities.

We now uplift the solution \eqref{metn4iia2}-\eqref{fluxn4iia2}
to M-theory. We get the M-theory background with metric
\begin{equation}
\label{metn4M2}
ds^2 = - \left( dt + \beta \sum_{i,j=1}^{8} J_{ij} x^i dx^j \right)^2
+ \left( du - \beta \sin \gamma \sum_{i,j=1}^{8} J_{ij} x^i dx^j \right)^2 
+ \sum_{i=1}^9 (dx^i)^2. 
\end{equation}
and four-form fluxes
\begin{eqnarray}
\label{fluxn4M2}
& F_{0129} = F_{0349} = F_{0569} = - F_{0789} = 2 \beta \sin \gamma 
\spa F_{1234} = F_{5678} = 4 \beta \cos \gamma, & \nn \\
& F_{u129} = F_{u349} = F_{u569} = - F_{u789} 
= 2 \beta. & 
\end{eqnarray}
where we denote $y=x^9$ and $u$ is the eleventh direction.
This solution has many common features with the 
$n=2$ \godel Universe/pp-wave mixture \eqref{metn2mix}-\eqref{fluxn2mix}.
For $\gamma=0$ it reduces to the
M-theory \godel Universe \eqref{metn4M1}-\eqref{fluxn4M1}.
For $\gamma=\pi/2$
it becomes an M-theory pp-wave with 22 supersymmetries
which was found in \cite{Gauntlett:2002cs}.
Thus, the solution interpolates between a \godel Universe and a
pp-wave, just as \eqref{metn2mix}-\eqref{fluxn2mix} does.
The structure of the metrics and fluxes are clearly also of the same
form.

The solution \eqref{metn4M2}-\eqref{fluxn4M2} has in general 
12 supersymmetries while it clearly has 20 supersymmetries 
when $\gamma = 0$ and 22 supersymmetries when $\gamma = \pi/2$
(see Appendix \ref{appsusy} for a proof of this).

Finally we note that 
if we set $\gamma = \pi/2$ we can illustrate that we do not have
any equivalence
between the T and S-dualities of Sections \ref{godelbyT} and \ref{godelbyS},
in the sense of the duality chain depicted in 
Figure \ref{figTS}, for $n=4$. For $\gamma = \pi/2$ 
\eqref{metpp28}-\eqref{fluxpp28} describes a type IIB pp-wave with
28 supersymmetries. The T-duality of Section \ref{godelbyT}
gives the type IIA \godel Universe
\eqref{metn4iia2}-\eqref{fluxn4iia2} with 20 supersymmetries.
Uplifting this to M-theory by the S-duality of Section \ref{godelbyS} 
gives an M-theory pp-wave with 22 supersymmetries. Thus, clearly 
the two pp-wave solutions related to the type IIA \godel Universe
cannot be related via any trivial dualities since they do not have the
same amount of supersymmetry.

\subsection{Solutions with $n=1$ and $n=3$}
\label{case1and3}

\subsubsection*{$n=1$ \godel Universes}

To find an $n=1$ \godel Universe, we start with the type IIB
pp-wave solution
\begin{equation}
\label{metppd5}
ds^2 = -2 dx^+ dx^- - \beta^2 \sum_{i=1}^2 (\tilde{x}^i)^2 (dx^+)^2
+ \sum_{i=1}^2 (d\tilde{x}^i)^2 + \sum_{i=3}^8 (dx^i)^2,
\end{equation}
\begin{equation}
\label{fluxppd5}
F^{(3)} = 2\beta dx^+ d\tilde{x}^1 d\tilde{x}^2,
\end{equation}
which is the Penrose limit of a D5-brane.
This pp-wave has 16 supersymmetries and is S-dual to the Nappi-Witten
model \cite{Nappi:1993ie,Russo:2002rq}, which is the 
Penrose limit of an NS5-brane \cite{Gomis:2002km}.

 We then do the T-duality of
Section \ref{godelbyT} with $a_1 = 1$. 
This gives the type IIA \godel Universe
\begin{equation}
\label{metgodeld5}
ds^2 = - \Big[ dt + \beta (x^1 dx^2 - x^2 dx^1) \Big]^2  
+ \sum_{i=1}^8 ( dx^i )^2  + dy^2,
\end{equation}
\begin{equation}
\label{fluxgodeld5}
H_{12y} = - 2\beta \spa
F_{12} = - 2\beta \spa
F_{012y} = 2\beta.
\end{equation}
Clearly this becomes a pp-wave when uplifted to M-theory, it in fact
becomes a pp-wave which is directly related to the pp-wave of 
\eqref{metppd5}-\eqref{fluxppd5}.
If we instead T-dualize the above type IIA solution along $x^3$ and $x^4$  
then we can trivially uplift this to M-theory and
we get the M-theory \godel Universe
\begin{equation}
\label{metgodeln1}
ds^2 = - \left[ dt + \beta (x^1 dx^2 - x^2 dx^1) \right]^2  
+ \sum_{i=1}^{\sharp} ( dx^i )^2  
\spa
F_{1234} = F_{5678} = - F_{129 \sharp} = 2\beta,
\end{equation}
where we put $y=x^9$ and $x^\sharp$ is the eleventh direction. 
The $G_3 \times \R^8$ \godel Universe 
\eqref{metgodeln1} has 8 supersymmetries.
We can write the solution as
\begin{equation}
(1,0,0,0,0) \spa (2,0,0,-2,0,0,0,2,0,0),
\end{equation}
in the notation of Section \ref{directgodel}.

\subsubsection*{$n=3$ \godel Universes}

Using the general formulas in \cite{Gauntlett:2002cs} we can construct the 
following M-theory pp-wave background with 20 supersymmetries
\begin{equation}
\label{metn3pp}
ds^2 = - 2dx^+ dx^- - \beta^2 \left( 
4 \Big[ (\tilde{x}^1)^2 + (\tilde{x}^2)^2 \Big]
+ \sum_{i=3}^6 (\tilde{x}^i)^2 \right) (dx^+)^2
+ \sum_{i=1}^6 (d\tilde{x}^i)^2 + \sum_{i=7}^9 (dx^i)^2,
\end{equation}
\begin{equation}
\label{fluxn3pp}
F_{(4)} = \beta dx^+ \left(
4 d\tilde{x}^1 d\tilde{x}^2 dx^7
+ 2 d\tilde{x}^3 d\tilde{x}^4 dx^7
+ 2 d\tilde{x}^5 d\tilde{x}^6 dx^7 \right).
\end{equation}
If we perform an S-duality along $x^9$ and the T-duality in
Section \ref{godelbyT} with $a_1 = 2$ and $a_2 = a_3 = 1$
we get the type IIB \godel Universe
\begin{equation}
\label{metn3iib}
ds^2 = - \left( dt + \beta \sum_{i,j=1}^6 J_{ij} x^i dx^j \right)^2  
+ \sum_{i=1}^8 ( dx^i )^2  + dy^2,
\end{equation}
\begin{eqnarray}
\label{fluxn3iib}
& H_{12y} = - 4\beta \spa H_{34y} = H_{56y} = - 2\beta \spa
F_{127} = - 4\beta \spa F_{347} = F_{567} = - 2\beta, & \nn \\
& F_{0127y} = - F_{34568} = 4\beta \spa
F_{0347y} = F_{0567y} = - F_{12568} = - F_{12348} = 2\beta, & 
\end{eqnarray}
with $J_{12} = 2$ and $J_{34} = J_{56} = 1$.
We now do a T-duality along $x^8$ and then we uplift the solution
to M-theory. This gives the M-theory background
\begin{equation}
\label{metn3M}
ds^2 = - \left( dt + \beta \sum_{i,j=1}^6 J_{ij} x^i dx^j \right)^2  
+ \sum_{i=1}^{\sharp} ( dx^i )^2, 
\end{equation}
\begin{eqnarray}
\label{fluxn3M}
& F_{3456} = F_{1278} = F_{129\sharp} = - 4\beta & \nn \\
& F_{1256} = F_{1234} = F_{3478} = F_{5678} = F_{349\sharp}
= F_{569\sharp} = - 2\beta, & 
\end{eqnarray}
with $y=x^9$ and $x^\sharp$ the eleventh direction.
This $G_7 \times \R^4$ \godel universe in M-theory preserves 
14 supersymmetries. We can write the solution as
\begin{equation}
(2,1,1,0,0) \spa (-2,-2,-4,-4,-4,-2,-2,-2,-2,0).
\end{equation}

\subsection{Solutions with $n=5$}
\label{case5}

We consider here M-theory \godel Universes with $n=5$. 
These backgrounds cannot be related to pp-waves in the same way
as for the $n \leq 4$ \godel Universes. To find these backgrounds
we have therefore instead used the methods of Section \ref{directgodel}.
The basic method used was to search through the possible $c_i$'s
with $c_i, a_{ij} \in \Z$. However, using these results as
input we have been able to find an $n=5$
solution with 18 supersymmetries and moreover a quite general family of 
solutions with at least 16 supersymmetries.

Consider the following \godel Universe background
\begin{equation}
\label{G11}
(2,1,1,1,1) \spa (-6,2,2,2,0,0,0,4,4,4),
\end{equation}
This is an $n=5$ \godel Universe with 18 supersymmetries.
We denote it $G_{11}$. We believe that this is the only possible
$n=5$ \godel Universe
with 18 supersymmetries, up to symmetries of the 
ansatz, and that it is not possible to
find an $n=5$ \godel Universe with more supersymmetry.
The metric and four-form flux are
\begin{equation}
\label{metn5}
ds^2 = - \left( dt + \beta \sum_{i,j=1}^{\sharp} J_{ij} x^i dx^j \right)^2
+ \sum_{i=1}^\sharp (dx^i)^2, 
\end{equation}
\begin{equation}
\label{fluxn5}
F_{1234} = - 6\beta \spa
F_{1256} = F_{1278} = F_{129\sharp} = 2\beta \spa
F_{5678} = F_{569\sharp} = F_{78\sharp} = 4\beta,
\end{equation}
with $J_{12} = 2$ and $J_{34} = J_{56} = J_{78} = J_{9\sharp} = 1$. 
In Appendix \ref{appsusy} we give the explicit expressions for the
Killing spinors of this background.

The $G_{11}$ solution \eqref{G11} is actually part of a family
of solutions
\begin{equation}
\label{almost}
(k+1,k,k,k,1) \spa (-4k-2,2,2,2k,0,0,2k-2,4k,2k+2,2k+2).
\end{equation}
This solution preserves at least 16 supersymmetries for any $k \in \R$.
We see that it reduces to \eqref{G11} for $k=1$.
Using the symmetries of the ansatz of Section \ref{directgodel}
we can in fact transform \eqref{almost} into the following 
general solution
\begin{equation}
\label{generalcase}
(p,p,p,p-q,sq) \ , \ (-4 s p, 0 , 2 s q , 2 p - 2 q , 0 , 2 s q , 2 p - 2 q
, 4 s p - 2 s q , 2 p + 2 q,2p), 
\end{equation}
with $p,q \in \R$ and $s = \pm 1$.
The general solution \eqref{generalcase} preserves at least
16 supersymmetries.
In fact all the $n=5$ solutions with at least
16 supersymmetries that we have found can fit into \eqref{generalcase}.
It is therefore natural to conjecture that 
any M-theory $n=5$ \godel Universe with at least 16 supersymmetries is a
special case of \eqref{generalcase}. We do not have a proof of this
at present.

We can also consider the special values of $p$ and $q$ which give
more than 16 supersymmetries for \eqref{generalcase}.
For $p=q=0$ we have $n=0$ and 32 supersymmetries, for
$p=q$ or $p=0$ we have $n=2$ and 20 supersymmetries, 
for $q=0$ we have $n=4$ and 20 supersymmetries 
and finally for $q=2p$ or $p=-q$ we have $n=5$ and 18 supersymmetries.

However, not all solutions of the ansatz
\eqref{metgodel5}-\eqref{fsansatz} which have at least
16 supersymmetries are special cases of \eqref{generalcase}. 
A counter-example is the $n=4$ \godel Universe
\begin{equation}
(2,2,1,1,0) \spa (0,-6,2,4,-2,6,4,0,2,2),
\end{equation}
which preserves 16 supersymmetries. 

We have also found several $n=5$ \godel Universe solutions with
less than 16 supersymmetries. We have studied all solutions
of the ansatz \eqref{metgodel5}-\eqref{fsansatz}
with $c_i, a_{ij} \in \Z$ and $1 \leq c_i \leq 5$. Of these we have found
one solution with 18 supersymmetries, being \eqref{G11},
several solutions with 16 supersymmetries all of which are special cases 
of \eqref{generalcase}, 
several solutions with 12 and 14 supersymmetries
and one solution without supersymmetry. Altogether we found 43 solutions.
We show here a list of five of these solutions.
\begin{displaymath}
\begin{array}{rcl}
\mbox{14 susy  } & \ \ (3,3,2,1,1) & \ \ \ \ (0,-10,4,4,-2,8,8,2,2,4), \\
\mbox{12 susy  } & \ \ (4,1,1,1,1) & \ \ \ \
 (-6,-6,-6,-6,-4,-4,-4,-4,-4,-4), \\
\mbox{12 susy  } & \ \ (5,2,1,1,1) & \ \ \ \
 (-6,-8,-8,-8,-6,-6,-6,-4,-4,-4), \\
\mbox{14 susy  } & \ \ (5,5,4,2,2) & \ \ \ \ (0,-18,6,6,-2,14,14,4,4,8). \\
\mbox{no susy  } & \ \ (5,5,4,2,2) & \ \ \ \ 
(-16,6,-10,2,6,2,-10,-12,-12,8). 
\end{array}
\end{displaymath}
%

\section{String Theory on \godel Universes}

In this section we compute the string spectrum
for two \godel Universe background of type IIA string theory. 
Our method is to consider the corresponding type IIB pp-wave background,
compactified on a circle of radius $R$,
and quantize the string theory in the light-cone gauge. 
Then we take the limit $R\to 0$ and obtain
the spectrum in the type IIA \godel Universe. 
We consider mainly the $n=4$ \godel Universe 
\eqref{metn4iia1}-\eqref{fluxn4iia1}
 T-dual to the maximally
supersymmetric pp-wave \eqref{metmaxpp}-\eqref{fluxmaxpp}, 
but also the $n=2$ \godel Universe
\eqref{G5IIA}-\eqref{fluxG5IIA} T-dual to the pp-wave
\eqref{metppintD3}-\eqref{fluxppintD3} is briefly considered.

\subsection{String Spectrum on compactified pp-wave}

In this section we 
compute the string spectrum in the compactified maximally 
supersymmetric type IIB pp-wave background
\eqref{metmaxpp}-\eqref{fluxmaxpp}
in the coordinate system of \eqref{aftercoII}.
Since we are only interested in the
supersymmetric case we choose $s=-1$.

We consider first the bosonic fields in the world-sheet theory
alone.
The light-cone gauge is defined by (see e.g.
\cite{Russo:1995tj,Russo:2002rq})
\ba
X^{+}(\tau,\sigma)=2\ap' p^+\tau+2wR\sigma \ \ \ (0\leq \sigma \leq \pi),
\ea
where $w$ is the winding number along the compactified circle. The momentum
is quantized as follows
\ba
p^+ = E+\f{m}{R},\ \ p^- = \f{E}{2}-\f{m}{2R}
\ \ \ (m\in {\bf Z}).
\ea
with $p^+ = -p_-$ and $p^- = - p_+$.
The world-sheet action in the light-cone gauge is given by
\ba
S=\f{1}{\pi\ap'}\int d\tau d\sigma \left[\di_{+}X^i\di_{-}X^i
-\beta \ap 'p^+ J_{ij}X^i\di_{\tau}X^j +\beta wR J_{ij}X^i\di_{\sigma}X^j
\right],
\ea
where we defined $\di_{\pm}=(\di_{\tau}\pm \di_{\sigma})/2$.
It is also useful to employ the complex fields
\ba
Z^{1}=X^1+iX^2,\ \ Z^{2}=X^3+iX^4, \ \ Z^{3}=X^5+iX^6, \ \ Z^{4}=X^7-iX^8.
\ea
The EOMs is given by
\ba
(\di_{\tau}^2-\di_{\sigma}^2) Z^a=4i\beta\ap' p^+ \di_{\tau}Z^a
-4i\beta wR \di_{\sigma}Z^a.  \ \ \ (a=1,2,3,4)
\ea
If we perform the field redefinition
\ba
Z^a=Z^a_0 e^{2i\beta p^+\tau +2i\beta^a wR\sigma}
(=Z^a_{0}e^{i\beta X^+}),
\ea
then we can write the EOMs as
\ba
(\di_{\tau}^2-\di_{\sigma}^2) Z^a_0+4f^2 Z^a_0=0,
\ea
where we defined
\ba
f=\s{(\beta\ap' p^+)^2-(\beta wR)^2}.
\ea
The boundary condition for $Z^a_0$ is given by
\ba
Z^a_0(\tau,\sigma+\pi)=e^{-2i\beta wR\pi}Z^a_0(\tau,\sigma). \label{tbc}
\ea
The fields $Z^a_0(\tau,\sigma)$ thus obey the same EOMs as in the
usual maximally supersymmetric IIB pp-wave.
However, it is important to notice that they
obey the twisted boundary condition \eqref{tbc}.
The mode expansion is given by ($\delta=\beta wR$)
\ba
Z^a_0&=&
i\s{\f{\ap'}{2}}\sum_{n\in {\bf Z}}
\left(\f{\ap^a_{n+\delta}}{\omega^+_{n}}
e^{-2i\omega^+_{n}\tau-2i(n+\delta)\sigma}+
\f{\ti{\ap}^a_{n-\delta}}{\omega^-_{n}}e^{-2i
\omega^-_{n}\tau+2i(n-\delta)\sigma}\right),\\
\bar{Z}^a_0&=&
i\s{\f{\ap'}{2}}\sum_{n\in {\bf Z}}
\left(\f{\bar{\ap}^a_{n-\delta}}{\omega^-_{n}}
e^{-2i\omega^-_{n}\tau-2i(n-\delta)\sigma}+
\f{\bar{\ti{\ap}}^a_{n+\delta}}{\omega^+_{n}}
e^{-2i\omega^+_{n}\tau+2i(n+\delta)\sigma}\right),
\ea
where we defined
\begin{equation}
\omega^+_{n} = \left\{ \begin{array}{ll} 
\s{(n + \delta)^2+f^2} & \mbox{ for } n \geq -\delta \\
-\s{(n + \delta)^2+f^2} & \mbox{ for } n < -\delta 
\end{array} \right.\ \ ,
\end{equation}
\begin{equation}
\omega^-_{n} = \left\{ \begin{array}{ll} 
\s{(n - \delta)^2+f^2} & \mbox{ for } n > \delta \\
-\s{(n - \delta)^2+f^2} & \mbox{ for } n \leq \delta 
\end{array} \right.\ \ .
\end{equation}

We consider now the light-cone gauge quantization of the string theory.
The quantization of the oscillators is again the same as
in the maximally supersymmetric IIB pp-wave with respect to $Z^a_0$ fields,
\begin{equation}
[\ap^a_{n+\delta},\bar{\ap}^b_{m-\delta}]
=2\omega^+_{n}\delta_{n+m,0}\delta_{a,b} \spa
[\ti{\ap}^a_{n-\delta},\ti{\bar{\ap}}^b_{m+\delta}]
=2\omega^-_{n}\delta_{n+m,0}\delta_{a,b} \ \ .
\end{equation}
We now compute the spectrum of the string theory by imposing the 
Virasoro constraints
$T_{++}=T_{--}=0$. Before we computing the spectrum, let us note that
\ba
\label{thepm}
p^-=-\f{\de S}{\de \dot{X}^+}=
\f{1}{4\pi\ap'}\int_{0}^{\pi}d\sigma \left[2\dot{X}^-
+\beta J_{ij}X^i\dot{X^j}\right] \ \ .
\ea
{}From this we can find $X^-=2\ap'p^-\tau-wR\sigma+\ddd$.
The condition $T_{++}+T_{--}=0$ leads to 
\begin{equation}
E^2-\left( \f{m}{R} \right)^2 - \left( \f{wR}{\ap'} \right)^2
= \frac{2}{\alpha'} \sum_{n \in \Z} \left(
N^{+}_{n}\s{(n+ \delta)^2+f^2}+N^{-}_{n}\s{(n-\delta)^2+f^2} \right)
+ 2\beta p^+ {\bf J}\ ,
\label{sp1}
\end{equation}
with the angular momentum operator
\begin{equation}
{\bf J} = \sum_{n \in \Z} \left( N^+_n - N^-_n \right) \ ,
\end{equation}
and the number operators
\begin{equation}
N^+_n = \left\{  \begin{array}{ll} 
\sum_{a=1}^4 \frac{1}{2\omega^+_{n}} 
\tilde{\alpha}_{-n-\delta}^a \tilde{\bar{\alpha}}_{n+\delta}^a 
& \mbox{ for } n \geq -\delta \\
\sum_{a=1}^4 \frac{1}{2\omega^-_{-n}} 
\alpha_{n+\delta}^a \bar{\alpha}_{-n-\delta}^a 
& \mbox{ for } n < -\delta
\end{array} \right. \ \ ,
\end{equation}
\begin{equation}
N^-_n = \left\{  \begin{array}{ll} 
\sum_{a=1}^4 \frac{1}{2\omega^-_{n}} 
\tilde{\bar{\alpha}}_{-n+\delta}^a \tilde{\alpha}_{n-\delta}^a   
& \mbox{ for } n > \delta \\
\sum_{a=1}^4 \frac{1}{2\omega^+_{-n}} 
\bar{\alpha}_{n-\delta}^a \alpha_{-n+\delta}^a 
& \mbox{ for } n \leq \delta
\end{array} \right. \ \ .
\end{equation}
In the above we ignored the zero-point energy 
since it is cancelled by the fermionic
one in the supersymmetric case $s=-1$. 

The other constraint $T_{++}-T_{--}=0$ leads to the level matching
\ba
\sum_{n=-\infty}^{\infty} n(N^+_{n}+N^-_{n})+mw=0.
\ea
Finally, we would like to note that 
the spectrum (\ref{sp1}) can also equivalently 
be found from the light-cone Hamiltonian
\ba
H&=&\f{1}{8(\ap')^2 p^+}\int^{\pi}_{0}d\sigma \left[\dot{Z}^a
\dot{\bar{Z}}^a+
Z^{a'}\bar{Z}^{a'}-2i\beta wR(Z^a\bar{Z}^{a'}- \bar{Z}^aZ^{a'}) \right],\no
&=&\f{1}{8(\ap')^2 p^+}\int^{\pi}_{0}
d\sigma \left[\dot{Z}_0^a \dot{\bar{Z}}_0^a+
Z_0^{a'} \bar{Z}_0^{a'}
+ 2i\beta \ap' p^+(Z_0^a\dot{\bar{Z}}_0^a-\bar{Z}_0^a\dot{Z}_0^a)
+4((\beta\ap'p^+)^2-(\beta Rw)^2)\right].\nonumber
\ea
Note that $H=p^-$ with $p^-$ given by \eqref{thepm} 
after imposing the Virasoro constraints.

We now turn to the contribution to the spectrum from world-sheet fermions.
We start with the general action
$S_{F}=\f{1}{\pi\ap'}\int d\tau d\sigma {\ca{L}}_F$ 
with \cite{Metsaev:2002re}
\ba
{\ca{L}}_F=i(\eta^{ab}\delta_{IJ}-\ep^{ab}\rho_{3IJ})\di_{a}X^{\mu}
\bar{\theta}^I\Gamma_{\mu}D_{b}\theta^{J},\label{lfa}
\ea
where $\theta^1$ and $\theta^2$ are Green-Schwarz fermions, each of
which is a Majorana-Weyl spinor in ten dimensions (sixteen components),
and 
where $\rho_{3}=\sigma_{3}$. The covariant derivative
is given by\footnote{Notice that the gamma matrix $\Gamma^\mu$ is
of the curved spacetime not of the local Minkowski frame. However, in the
discussion below we can neglect this difference as one can see explicitly.}
\ba
D_{a}\!=\!\di_{a}+\f{1}{4}\di_{a}X^{\ap}\left(\omega_{\mu\nu\ap}\Gamma^{\mu\nu}
-\f{1}{4\cdot 5!}F_{\mu\nu\rho\lambda\kappa}
\Gamma^{\mu\nu\rho\lambda\kappa}(i\sigma_2)\Gamma_{\ap}\right)\!,\label{lfb}
\ea
where the non-trivial components of the spin connection
are given by $\omega_{+ij}=\beta J_{ij}$.
After we impose
the light-cone gauge $\Gamma^{+}\theta^{1,2}=0$ for fermions,
we find the following Lagrangian for the fermions\footnote{To derive this
we use the properties of $\Gamma$ matrices
$\{\Gamma^\mu,\Gamma^\nu\}=2g^{\mu\nu},\ \ (\Gamma^{0})^{T}=-\Gamma^{0},
\ \ (\Gamma^{i})^{T}=\Gamma^{i}$. It turns out that only 
the component $\di_{a}X^{+}$ 
contributes in (\ref{lfa}) and (\ref{lfb}).}
\ba
{\ca{L}}_F&=&4i(\ap'p^+ -wR)\bar{\theta}^1\Gamma_{+}
\left(\di_{+}-i\beta(\ap' p^+ +wR)J\right)\theta^1\no
&&+4i(\ap'p^+ +wR)\bar{\theta}^2\Gamma_{+}
\left(\di_{-}-i\beta(\ap' p^+ -wR)J\right)\theta^2\no
&&-8i\beta\left((\ap'p^+)^2 -(wR)^2\right)
\bar{\theta}^1\Gamma^+\Gamma^{1234}\theta^2,
\ea
where $J$ denotes
the spin $J=\f{i}{4}J_{ij}\Gamma^{ij}$ of the field $\theta^{1,2}$.
After normalizing the fermions (both have eight components
as spinors in 8 dimensions) such that
$\theta^1=2\s{\ap'p^+ -wR}\ S^1$ 
and $\theta^2=2\s{\ap'p^+ +wR}\ S^2$, we obtain
\ba
{\ca{L}}_F&=&i\Bigl(S^1
\left(\di_{+}-i\beta(\ap' p^+ +wR)J\right)
S^1+
S^2\left(\di_{-}-i\beta(\ap' p^+ -wR)J\right)S^2\no
&&-2\beta\s{(\ap'p^+)^2 -(wR)^2}S^1\gamma^{1234}S^2\Bigr),
\ea
where we define the Gamma-matrices 
in eight dimensions by $\gamma^i\  (i=1,2,\ddd,8)$.
Defining the fermionic fields $\ca{S}^1$ and $\ca{S}^2$ 
as follows
\ba
S^1(\tau,\sigma)
&=&\exp\left(2i\beta J(\ap'p^+\tau+wR\sigma)\right)\ca{S}^1(\tau,\sigma),\no
S^2(\tau,\sigma)
&=&\exp\left(2i\beta J(\ap'p^+\tau-wR\sigma)\right)\ca{S}^2(\tau,\sigma),
\ea
then the action is equivalent to that of 
the maximally supersymmetric IIB pp-wave. Notice that,
as for the bosonic fields,
we have a twisted boundary condition
\ba
\ca{S}^{1,2}(\tau,\sigma+\pi)=e^{-2i\beta \pi JwR} \ca{S}^{1,2}(\tau,\sigma).
\label{ftbc}
\ea
To diagonalize the matrix $J$ we consider the eigenvalues of
$(i\Gamma_{12},i\Gamma_{34},i\Gamma_{56}, i\Gamma_{78})$ since the
spin $J$ is given by 
$J=\f{i}{2}(\Gamma_{12}+\Gamma_{34}+\Gamma_{56}-\Gamma_{78})$. 
 Then
the eight possible cases can be divided into four with $J=1$, i.e.  
$(+,+,+,+),(-,+,+,-),(+,-,+,-),(+,+,-,-)$, and four with $J=-1$, i.e. 
$(-,+,-,+),(-,-,+,+),(+,-,-,+),(-,-,-,-)$. Here we used the
constraint $\gamma^{1234}S^{1,2}=\gamma^{5678}S^{1,2}$. In this way we have
found that the fermionic fields $\ca{S}^1$ and $\ca{S}^2$ have 
the same twisted boundary condition
(\ref{ftbc}) as the bosonic ones. This is due to 
supersymmetry and leads to
vanishing zero-point energy.
Since complex conjugation of a spinor change the sign
of its eigenvalue under $J$,
we can conveniently use the complex fermion fields
$S^{Ii}$ with the charge $J=-1$ and $\bar{S}^{Ii}\ \ (I=1,2,\ i=1,2,3,4)$ 
with $J=1$ below.

We can now compute the fermionic part of the light-cone Hamiltonian
\ba
H&=&\f{i}{32\pi(\ap')^2 p^+}\int_0^{\pi} d\sigma\left[
\bar{S}^1 \dot{S}^1 + S^1 \dot{\bar{S}}^1
+ \bar{S}^2 \dot{S}^2 + S^2 \dot{\bar{S}}^2 \right],\label{hf}\\
&=&\f{i}{32\pi(\ap')^2 p^+}\int_0^{\pi} d\sigma
\left[\bar{\ca{S}}^1 \dot{\ca{S}}^1
+\ca{S}^1\dot{\bar{\ca{S}}}^1+
\bar{\ca{S}}^2 \dot{\ca{S}}^2
+\ca{S}^2\dot{\bar{\ca{S}}}^2
+2i\beta\ap' p^+ J (\bar{\ca{S}}^1\ca{S}^1+\ca{S}^2 \bar{\ca{S}}^2)
\right].\nonumber
\ea
We expand the (complex) fermionic field as follows ($i=1,2,3,4$)
\ba
S^{1i}(\sigma,\tau)&=&\s{\ap'} \sum_{n\in \Z} \Biggl[
\f{f}{\s{f^2+(\omega^+_{n}-f)^2}}
 e^{-2i\omega^+_{n}\tau+2i (n+\delta)\sigma}S^i_{n+\delta}\no
&&+\f{i(\omega^-_{n}-(n-\delta))}
{\s{f^2+(\omega^-_{n}-f)^2}}
e^{-2i\omega^-_{n}\tau -2i (n-\delta)\sigma}
(\Pi \ti{S}^i_{n-\delta})\Biggr],\no
S^{2i}(\sigma,\tau)&=&\s{\ap'} \sum_{n\in \Z} \Biggl[\f{f}
{\s{f^2+(\omega^-_{n}-f)^2}}
e^{-2i\omega^-_{n}\tau-2i (n-\delta)\sigma}\ti{S}^i_{n-\delta}\no
&&-\f{i(\omega^+_{n}-(n+\delta))}
{\s{f^2+(\omega^+_{n}-f)^2}}
e^{-2i\omega^+_{n}\tau +2i (n+\delta)\sigma}(\Pi 
{S}^i_{n+\delta})\Biggr], 
\ea
where $\Pi = \gamma^{1234}$ and $\delta=\beta w R$ as above.
The quantization of oscillators is given by
\begin{equation}
\{S^i_{n+\delta},\bar{S}^j_{m-\delta}\}=2\delta_{n+m,0}\delta_{ij}
\spa
\{\tilde{S}^i_{n-\delta},\ti{\bar{S}}^j_{m+\delta}\}
=2\delta_{n+m,0}\delta_{ij} \ .
\end{equation}
Finally we can write the Hamiltonian (\ref{hf}) as follows
\begin{equation}
\label{fermham}
H = \f{1}{\ap'p^+} \sum_{n \in \Z} \left(
F^+_n \sqrt{(n+\delta)^2 + f^2 } 
+ F^-_n \sqrt{(n-\delta)^2 + f^2 } \right)
+ \beta \sum_{n \in \Z} ( F^+_n - F^-_n ) \ ,
\end{equation}
with the fermionic number operators given by
\begin{equation}
F^+_n = \left\{  \begin{array}{ll} 
\frac{1}{2} \sum_{i=1}^4 
\tilde{S}_{-n-\delta}^i \tilde{\bar{S}}_{n+\delta}^i 
& \mbox{ for } n \geq -\delta \\
\frac{1}{2} \sum_{i=1}^4 
S_{n+\delta}^i \bar{S}_{-n-\delta}^i 
& \mbox{ for } n < -\delta
\end{array} \right. \ \ ,
\end{equation}
\begin{equation}
F^-_n = \left\{  \begin{array}{ll} 
\frac{1}{2} \sum_{i=1}^4 
\tilde{\bar{S}}_{-n+\delta}^i \tilde{S}_{n-\delta}^i  
& \mbox{ for } n > \delta \\
\frac{1}{2} \sum_{i=1}^4 
\bar{S}_{n-\delta}^i S_{-n+\delta}^i 
& \mbox{ for } n \leq \delta
\end{array} \right. \ \ .
\end{equation}
{}From \eqref{sp1} and \eqref{fermham} we see that the 
total spectrum of the string theory on
the compactified maximally 
supersymmetric pp-wave background \eqref{metmaxpp}-\eqref{fluxmaxpp} is
\begin{eqnarray}
&& E^2-\left( \f{m}{R} \right)^2 - \left( \f{wR}{\ap'} \right)^2
= 2\beta p^+ \sum_{n \in \Z} 
\left( N^{+}_{n}+F^{+}_{n} - N^{-}_{n}- F^{-}_{n} \right) 
 \nn \\
&& + \frac{2}{\alpha'} \sum_{n \in \Z} \left(
(N^{+}_{n}+F^{+}_{n})\s{(n+ \delta)^2+f^2}
+(N^{-}_{n}+F^{-}_{n})\s{(n-\delta)^2+f^2} \right)  \ ,
\label{totspec}
\end{eqnarray}
with level matching condition
\begin{equation}
\sum_{n \in \Z} n(N^+_{n}+N^-_{n} + F^+_n + F^-_n)+mw=0.
\end{equation}
We see that the supersymmetry is manifest in these expressions.

\subsection{String Spectrum on \godel Universe}

We consider now the T-dual background in type IIA string theory 
and take the limit $R\to 0$.
Then we find that the string 
spectrum on the type IIA \godel Universe with 20 supersymmetries 
\eqref{metn4iia1}-\eqref{fluxn4iia1}
is given by
\begin{eqnarray}
\label{spcg}
&& E^2-p_{y}^2 = 
\f{2}{\ap'} \sum_{n \in \Z} \left[ (N^{+}_{n} + F^+_n) 
\s{(n + \ap'\beta p_{y})^2 +\beta^2 {\ap'}^2 (E^2-p_{y}^2)}
\right. \\ 
&& \left.
+ (N^{-}_{n} + F^-_n) \s{(n - \ap'\beta p_{y})^2
+\beta^2 {\ap'}^2 (E^2-p_{y}^2)} \right]
+ 2\beta E \sum_{n \in \Z} 
\left( N^{+}_{n}+F^{+}_{n} - N^{-}_{n}- F^{-}_{n} \right)\ , \nn
\end{eqnarray}
along with the level matching condition
\begin{equation}
\sum_{n \in \Z} n(N^+_{n}+N^-_{n} + F^+_n + F^-_n)=0.
\end{equation}

As a check of \eqref{spcg} we compare now the spectrum of the 
bosonic zero-modes
to the spectrum computed from supergravity.
If we restrict to the zero-modes, then \eqref{spcg} becomes
\ba
E^2-p_{y}^2
=2\beta E({\bf N}+{\bf J}+\ep_0),\label{massless}
\ea
where we set $N^+_{0}+N^-_{0} + F^+_0 + F^-_0={\bf N}+\ep_0$ and
$\ep_0(\geq 0)$ represents the fermionic 
zero-point energy (see e.g. \cite{Metsaev:2002re} 
for the detailed analysis of the zero-point energy in the pp-wave).

Let us for example consider the dilaton field 
$\varphi$ ($\ep_0=4$) in the \godel background and compare the
above result with that obtained from the low energy supergravity analysis.
The EOM is given by the massless Klein-Gordon equation
\ba
\square \varphi=\left(\f{\di^2}{\di y^2}
-\f{\di^2}{\di t^2}\right)\varphi
+\sum_{a=1}^4 \left(\vartriangle_{a} +
\beta^2 \rho_a^2
\f{\di^2 }{\di t^2}
-2\beta\f{\di^2}{\di t \di \phi_a}\right)\varphi=0,
\ea
where we denote the Laplacian by $\vartriangle_{a}=
\f{1}{\rho_a}\f{\di }{\di \rho_a}(\rho_a \f{\di}{\di \rho_a})+
\f{1}{\rho_a^2}\f{\di^2}{\di \phi_{a2}}$. 
We assume the form
\ba
\varphi=e^{-iEt+i{\bf J_a}\phi_a+ip_{y}y}f(\rho),
\ea
and then we obtain the harmonic system
\ba
\sum_{a=1}^4 \left(-\vartriangle_{a}+(\beta E)^2 \rho_a^2
\right)f(\rho)
=\left(E^2-p_{y}^2- 2\beta E {\bf J}\right)f(\rho), 
\ea
where ${\bf J}=\sum_{a=1}^4{\bf J}_a$. By using standard 
results of harmonic oscillators we find the spectrum
\ba
E^2-p_{y}^2= 2\beta E ({\bf N}+{\bf J}+4),\label{specte}
\ea
where ${\bf N}=n_1+n_2+\ddd +n_8$ is the familiar quantum number
of eight harmonic oscillators. This result exactly matches with the string
theory result (\ref{massless}).

Clearly, the equation for the 
spectrum \eqref{spcg} is of a rather intricate nature
since it has energy both on the left- and right-hand side.
In order to gain a better understanding of \eqref{spcg} we consider
the special case where $p_y = 0$. We then get
\begin{equation}
\frac{E^2}{2\beta} = 
\sum_{n \in \Z} \s{E^2 + \left( \frac{n}{\beta \ap'} \right)^2 } 
\Big( \ca{N}^+_n + \ca{N}^-_n  \Big) 
+ E \sum_{n \in \Z} 
\left( \ca{N}^+_n - \ca{N}^-_n \right)\ ,
\end{equation}
where we define
\begin{equation}
\ca{N}^+_n = N^+_n + F^+_n \spa
\ca{N}^-_n = N^-_n + F^-_n \ .
\end{equation}
If we consider the case where the string only has low-lying
string modes \( |n| \ll \beta \alpha' E \) the spectrum can
be written
\begin{equation}
E^3 = 4 \beta E^2 \sum_{n \in \Z} \ca{N}^+_n
+ \frac{1}{\beta (\alpha')^2 } \sum_{n \in \Z} n^2 
\left( \ca{N}^+_n + \ca{N}^-_n \right)
\end{equation}
If we can further neglect the first term, for example by considering
$\ca{N}^+_n = 0$, we see that we get a very unconventional string spectrum.

If we instead consider the case where the string only has
high excitations \( |n| \gg \beta \alpha' E \) the spectrum can
be written as
\begin{equation}
E^2 = \frac{2}{\alpha'} \sum_{n \in \Z} 
|n| \left( \ca{N}^+_n + \ca{N}^-_n \right)
+ 2\beta E \sum_{n \in \Z} \left( \ca{N}^+_n - \ca{N}^-_n \right)
+ \beta^2 \alpha' E^2 \sum_{n \in \Z} 
\frac{1}{|n|} \left( \ca{N}^+_n + \ca{N}^-_n \right)\ ,
\end{equation}
where the last term is small compared to the first term.
If we throw away the last term we can write
\begin{equation}
E = \beta \sum_{n \in \Z} \left( \ca{N}^+_n - \ca{N}^-_n \right)
+ \sqrt{ \left[ \beta \sum_{n \in \Z} \left( \ca{N}^+_n - \ca{N}^-_n \right)
\right]^2 + \frac{2}{\alpha'} \sum_{n \in \Z} 
|n| \left( \ca{N}^+_n + \ca{N}^-_n \right) }
\end{equation}
We see that we regain the flat-space string spectrum if we can neglect
the angular momentum part. 

Finally, going back to the spectrum \eqref{spcg} with $p_y$, 
we note that if we take the $\beta^2 \alpha' \rightarrow \infty$ limit
we get the spectrum
\begin{equation}
E = \frac{1}{2} p_y + \sqrt{ \frac{1}{4} p_y^2
+ 4 \beta \sum_{n \in \Z} \ca{N}^+_n }
\label{betabig}
\end{equation}
The limit $\beta^2 \alpha' \rightarrow \infty$ means the geometry
become strongly curved as we can understand from (\ref{curvscal}). 
As we can see from \eqref{betabig}
this means the masses of the string states become negligible.
This is why the spectrum \eqref{betabig} effectively becomes
that of the supergravity modes \eqref{massless}.
This limit is obviously interesting since the radii of
the closed time-like curves would be able 
to become arbitrarily small in this limit
if the supergravity description was valid.
We also note that the spectrum \eqref{betabig} suggest that one can make
a string bit model similar to the one of 
\cite{Spradlin:2002ar,Kristjansen:2002bb,Constable:2002hw,Verlinde:2002ig} 
on the maximally supersymmetric pp-wave.

\subsection{$n=2$ \godel Universe in type IIA}

As the next case
we examine the string theory in the \godel background 
which is
T-dual to the supersymmetric pp-wave defined
by \eqref{metppintD3} and \eqref{fluxppintD3}. This background 
is given by (\ref{G5IIA}) and (\ref{fluxG5IIA}) and 
preserves
20 supersymmetries as we have seen above.

The computation of spectrum in the compactified pp-wave can be performed
as in the previous case.
For example, the fermionic part is given by
\ba
{\ca{L}}_F&=&i\Bigl(S^1
\left(\di_{+}-i\beta(\ap' p^+ +wR)J\right)
S^1+
S^2\left(\di_{-}-i\beta(\ap' p^+ -wR)J\right)S^2\no
&&-\beta\s{(\ap'p^+)^2 -(wR)^2}S^1(\Gamma^{1256}+\Gamma^{3456})
S^2\Bigr),
\ea
where the matrix $J$ is given by $J=\f{i}{4}\sum_{i,j=1}^4
J_{ij}\Gamma^{ij}$. It is useful to take the following 
linear combinations
for each of the 
two eight components spinors $S_a^1$ and $S_a^2\ \ (a=1,2,\ddd,8)$
\ba
&&S_1: (+,+,+,+), \ \ S_2: (+,+,-,-), \ \ S_3: (+,-,+,-),
\ \ S_4: (-,+,+,-), \no
&&\bar{S}_1:(-,-,-,-) , \ \ \bar{S}_2: (-,-,+,+) , \ \ \bar{S}_3:(-,+,-,+)
, \ \ \bar{S}_4:(+,-,-,+) , 
\ea
where we specified the eigenvalue $\pm 1$ of $(i\Gamma_{12},
i\Gamma_{34},i\Gamma_{56},i\Gamma_{78})$.
Then it is easy to see that
\ba
(\Gamma^{12}+\Gamma^{34})S_{3,4}=(\Gamma^{12}+\Gamma^{34})\bar{S}_{3,4}=0,
\ea
and thus these four spinors out of eight are massless, while 
the other four are
massive. The value of
$J$ is given by zero for $S_{3,4}, \bar{S}_{3,4}$ and $+1$ ($-1$) for
$S_{1,2}$ ($\bar{S}_{1,2}$). This charge distribution (or equally the
twisted boundary condition) is the same as in the
bosonic fields in the world-sheet theory and this is again due to the
supersymmetry in the compactified solution. 

The string spectrum is given by the previous formula (\ref{sp1})
with respect to the $Z^{1,2},S_{1,2}$ ($J=1$) 
and $\bar{Z}^{1,2},\bar{S}_{1,2}$ ($J=-1$), 
while we should set
$\beta=0$ with respect to
$Z^{3,4},S_{3,4}$ and $\bar{Z}^{3,4},\bar{S}_{3,4}$ ($J=0$)
excitations. The zero-energy does vanish
due to the remaining supersymmetry. The small radius limit leads to the \godel
Universe model and the spectrum is given by
\begin{eqnarray}
&& E^2-p_{y}^2 = 
\f{2}{\ap'} \sum_{n \in \Z} \left[ (N^{+}_{n} + F^+_n) 
\s{(n + \ap'\beta p_{y})^2 +\beta^2 {\ap'}^2 (E^2-p_{y}^2)}
\right. \nn \\ 
&& \left.
+ (N^{-}_{n} + F^-_n) \s{(n - \ap'\beta p_{y})^2
+\beta^2 {\ap'}^2 (E^2-p_{y}^2)} \right]
+ 2\beta E \sum_{n \in \Z} 
\left( N^{+}_{n}+F^{+}_{n} - N^{-}_{n}- F^{-}_{n} \right)
\nn \\ &&
+ \sum_{i=1}^4 p_i^2 
+ \f{2}{\ap'}\sum_{n=-\infty}^{\infty} |n| ( N_n + F_n ) \ , 
\end{eqnarray}
along with the level matching condition
\begin{equation}
\sum_{n \in \Z} n(N^+_{n}+N^-_{n} + N_n + F^+_n + F^-_n + F_n )=0.
\end{equation}
where $N^{\pm}_{n},F^{\pm}_{n}$ 
and $N_n,F_n$ counts the number of oscillators with
spin $J=\pm 1$ and $J=0$, respectively, and $p_i$ are the momenta in the four
extra transverse directions.
Obviously, other supersymmetric 
\godel Universe backgrounds of string theory can be treated similarly using
these methods.

\section{D-branes on \godel Universes}

In this section we consider the D-brane spectrum on \godel Universes 
which are T-dual to highly supersymmetric pp-waves. We obtain 
the D-brane spectrum by taking the T-duality transformation of that 
in the pp-wave. Below we investigate this issue from the viewpoint of 
both the boundary condition in the world-sheet theory and the classical
brane solutions in supergravity. We consider the cases of the \godel
Universes \eqref{metn4iia1}-\eqref{fluxn4iia1} 
and \eqref{G5IIA}, \eqref{fluxG5IIA2}.

\subsection{Boundary Conditions in String Theory}

We discuss here
the D-branes in the type IIA $n=4$ \godel Universe with 
20 supersymmetries,
defined by \eqref{metn4iia1} and \eqref{fluxn4iia1} with $s=-1$,
by applying T-duality to the known D-brane spectrum in the maximally 
supersymmetric type IIB pp-wave. We consider here
the D-branes on the pp-wave that have Neumann
boundary conditions in the $x^+$ and $x^-$ directions.
Then the 
half-BPS D-brane in the pp-wave background is given by 
the D$p$-branes ($p=3,5,7$) which have the boundary
condition $(+,-,2,0)$, $(+,-,3,1)$ and $(+,-,4,2)$ respectively%
\footnote{Here the symbol $(+,-,a,b)$ means Neumann boundary 
conditions for 
$x^+,x^-,\ti{x}^{i_1},...,\ti{x}^{i_a},\ti{x}^{j_1},...,\ti{x}^{j_b}\ \ 
(1\leq i_1,\ddd,i_a \leq 4,\ \ 5\leq j_1,\ddd,j_a \leq 8)$ following the
convention \cite{Skenderis:2002vf}.} 
\cite{Dabholkar:2002zc} 
(see also \cite{Billo:2002ff,Skenderis:2002vf}). The spectrum also 
includes the 1/4 BPS D1-brane $(+,-,0,0)$ \cite{Skenderis:2002vf}. Below 
we study the T-dual counterparts of these branes.

To investigate the D-brane spectrum we consider in the following 
the boundary conditions for D-branes
(see also \cite{Takayanagi:2001gu} for similar analysis 
in the exactly solvable model of magnetic universe \cite{Russo:1995tj}).  
For simplicity we discuss only the bosonic part of the boundary conditions.
Under the T-duality in the $y=\frac{1}{2}x^+ - x^-$ direction 
the world-sheet fields are transformed as follows.%
\footnote{For a general formula see e.g. \cite{Russo:1995tj} 
It is given by
\ba
\di\ti{y}&=&-B_{y\mu}\di x^{\mu}-g_{yy}(\di y+G_{y\mu}\di x^\mu), \no
\bar{\di}\ti{y}&=&-B_{y\mu}\bar{\di} x^{\mu}
+g_{yy}(\bar{\di} y+G_{y\mu}\bar{\di} x^\mu).
\ea}
\ba
\di_D y&=&\di_N \ti{y}+\beta(\rho_1^2\di_D\phi_1+\rho_2^2\di_D\phi_2
+\rho_3^2\di_D\phi_3-\rho_4^2\di_D\phi_4),\no
\di_N y&=&\di_D \ti{y}+\beta(\rho_1^2\di_N\phi_1+\rho_2^2\di_N\phi_2
+\rho_3^2\di_N\phi_3-\rho_4^2\di_N\phi_4),
\label{tdb}
\ea
where we defined $\di_{N}=\di+\bar{\di}=\di_{\sigma}$ and $\di_{D}=
\bar{\di}-\di=-i\di_{\tau}$. 
Here and in the rest of this section we call
the T-duality direction $y$ in the pp-wave background
and $\tilde{y}$ in the T-dual \godel Universe background.

In order to take the T-duality we assume the D-brane is placed away
from the origin at fixed $x^i\neq 0\ \ (i=1,2,\ddd,8)$. 
Thus, while the D-brane is at a fixed position in the $x^i$ coordinates
(and therefore in the $\rho_k$ and $\phi_k$ coordinates),
the position of the D-brane
is rotating as $\ti{x}^{2k-1}= \rho_k \cos( \phi_k - a_k \beta x^+)$ and 
$\ti{x}^{2k}= \rho_k \sin( \phi_k - a_k \beta x^+)\ \ (k=1,2,3,4)$
in the original coordinates of the pp-wave \eqref{metmaxpp}.
These shifted D-brane configurations can be obtained by a symmetry
transformation described in \cite{Skenderis:2002wx} and they
have the same amount of supersymmetry as before.

\subsubsection*{D0-branes from D1-branes}

Before the T-duality
a D1-brane on the pp-wave has the following boundary conditions
\ba
&&\di_{D}{\rho_{i}}=0,\ \ \di_{D}{\phi_{i}}=0,\ \  (i=1,2,3,4) 
\label{bcd11} \\
&&\di_{N}x^{+}=0,\label{bcd12} \\
&&\di_{N}x^-+\beta (\rho_1^2\di_N \phi_1+\rho_2^2\di_N \phi_2
+\rho_3^2 \di_N  \phi_3
-\rho_4^2 \di_N \phi_4)=0,\label{bcd13}
\ea
where \eqref{bcd13} comes from the consideration of symmetries
acting on boundary conditions \cite{Skenderis:2002wx}. 
After rewriting the above boundary conditions using (\ref{tdb}), we get 
\ba
&&\di_{N}t+\beta (\rho_1^2\di_N \phi_1+\rho_2^2\di_N \phi_2
+\rho_3^2 \di_N  \phi_3
-\rho_4^2 \di_N \phi_4)=0, \label{bcd1tt}\\
&&\di_{D}\ti{y}=0, \label{bcd1t}
\ea
with (\ref{bcd11}).  Finally let us compare this result with
the general boundary conditions%
\footnote{Here the index $\mu$ in the mixed Neumann condition
 should be taken along the world-volume direction.}
\ba
&&G_{\mu\nu}\di_{N}X^\nu+(B_{\mu\nu}+F_{\mu\nu})
 \di_{D}X^\nu|_{\di\Sigma}=0\ \ \
 (\mbox{mixed Neumann})\label{bcnn1} \\ &&\mbox{or}\ \ \
 \di_{D}f(X^\mu)|_{\di\Sigma}=0  \ \ \ 
(\mbox{Dirichlet}) \label{bcnn2}.
\ea
In conclusion the boundary conditions \eqref{bcd1tt} and \eqref{bcd1t}
shows that the D-brane obtained by T-duality is a D0-brane in the \godel
background. Indeed its world-volume $x^i=y=const.$ 
is equivalent to the time-like geodesic line.

\subsubsection*{D2 and D4-branes from D3-branes}

Let us consider a D3-brane whose world-volume is along 
$(x^+,x^-,\ti{x}^1,\ti{x}^2)$
in the pp-wave background. The computations are almost the same as above.
We obtain \eqref{bcd1t} again by T-duality (the other conditions are the same
 as before T-duality). Thus we get a D2-brane extended in the $x^1,x^2$ 
which means we have a ``small'' 
\godel Universe in the world-volume directions. 
 
A more interesting case is the D3-brane whose world-volume is given by
$(x^+,x^-,\ti{x}^1,\ti{x}^3)$. In this case we start with the following 
boundary conditions
in the pp-wave
\ba
&&\di_{N}{\rho_{1,2}}=0,\ \ \di_{D}{\rho_{3,4}}=0,\ 
\label{bcd31} \\
&&\di_{D}(\phi_1-\beta x^+)=\di_{D}(\phi_2-\beta x^+)=0,\ \ 
\di_{D}\phi_{3,4}=0, \\
&&\di_{N}x^{+}=0,\label{bcd32} \\
&&\di_{N}x^-+\beta (\rho_3^2 \di_N  \phi_3
-\rho_4^2 \di_N \phi_4)=0.\label{bcd33}
\ea
After we take T-duality, the result is given by (\ref{bcd1tt}) and
\ba
&&\di_{D}(\phi_1-\phi_2)=0,\\
&&\beta (\rho_1^2 \di_{N}\phi_1+\rho_2^2 \di_{N}\phi_2)+
\di_{D}\ti{y}=0,\\
&&\di_{N}\ti{y}+\di_{D}t+
\beta (\rho_1^2 \di_{D}\phi_1+\rho_2^2 \di_{D}\phi_2)-\f{1}{\beta}
\di_{D}\phi_1=0.
\ea
Comparing with (\ref{bcnn1}) and (\ref{bcnn2}), 
we find that the
resulting system represents a D4-brane whose world-volume coordinate 
is given by
$t,\ti{y},\rho_1,\rho_2$ and $\theta=(\phi_1+\phi_2)/2$ with the gauge flux
\ba
F_{t\ti{y}}=-1,\ \ F_{\ti{y}\theta}=-\f{1}{\beta}.\label{fluxb}
\ea

\subsubsection*{D6-branes from D5-branes}

We can assume that the world-volume of D5-brane is in the direction of
$(x^+,x^-,\ti{x}^1,\ti{x}^2,\ti{x}^3,\ti{x}^5)$. Then we find almost 
the same boundary conditions as for the D4-brane from the D3-brane,  
and the result is given by a D6-brane
extending in the direction 
$t,\ti{y},\rho_1,\phi_1,\rho_2,\rho_3$ and $\theta=(\phi_2+\phi_3)/2$
with the flux (\ref{fluxb}).

\subsubsection*{D6 and D8-branes from D7-branes}

If we consider the D7-brane extending
in the direction $(x^{+},x^{-},\ti{x}^1,\ti{x}^2,\ti{x}^3,\ti{x}^4,\ti{x}^5,\ti{x}^6)$, then we get
D6-brane (with no gauge flux) in the \godel model as in the same way as in
the D1-brane in pp-wave.

For the D7-brane whose world-volume is given by
$(x^{+},x^{-},\ti{x}^1,\ti{x}^2,\ti{x}^3,\ti{x}^4,\ti{x}^5,\ti{x}^7)$, we obtain
a D8-brane in the direction $t,\ti{y},
\rho_1,\rho_2,\rho_3,\rho_4,\phi_1,\phi_2$ and $\theta=(\phi_3+\phi_4)/2$
with the gauge flux (\ref{fluxb}) in the same way as when we got a
D4-brane from a D3-brane.

\subsubsection*{Summary of D-branes in the \godel model}

In this way we get D0,D2,D4,D6 and D8-branes in the \godel background.
In particular we have obtained the D-branes
with both electric and magnetic fluxes for D4,D6 and D8-branes. 
Apart from the D0-branes, 
the above list of D-branes have world-volumes 
which include closed time-like curves so that we have 
``small'' \godel Universes along their world-volumes.
It would be interesting to study the gauge theories on these
world-volumes.

\subsection{Classical Brane Solutions in Supergravity}

We now consider D-branes on \godel Universes from the 
viewpoint of the supergravity solutions. The example we discuss
here is that of a D4-brane in the $n=2$ \godel model 
given by \eqref{G5IIA}, \eqref{fluxG5IIA2} with 20 
supersymmetries. We find the supergravity solution for a smeared
D4-brane by performing 
T-duality transformations of a D5-brane solution in the pp-wave
\eqref{metppintD3}, \eqref{fluxppD1D5}. The D5-brane solution was 
found in \cite{Kumar:2002ps} (see also \cite{Singh:2002yd} 
and \cite{Alishahiha:2002rw} 
for related systems). Afterwards we consider the localized D4-brane.

The D5-brane solution on the pp-wave is
\ba
&&ds^2=h^{-\f12}\Bigl(-2dx^+ dx^- -\beta^2\sum_{i=1}^4(\ti{x}^i)^2
(dx^{+})^2+\sum_{i=1}^4{(d\ti{x}^i)^2}\Bigr)
+h^{\f12}\sum_{j=5}^8{(dx^j)^2},\no
&&h=1+\f{Ng_s\ap '}{r^2},\ \ \ e^{2\phi}=g^{-1},\ \ 
\ F_{+12}=F_{+34}=2\beta,\no
&& \ \ F_{ijk}=-\ep_{ijkl}\di_{l}h,\ (5\leq i,j,k,l \leq 8),
\ea
where $N$ is the number of D5-branes and also 
we defined $r=\s{\sum_{i=5}^8(\ti{x}^i)^2}$ .
If we simply take the T-duality as before 
we obtain the smeared D4-brane
solution in the \godel model. To localize this solution 
we have to replace the function
$h$ with 
\ba
f=1+\f{\pi Ng_s(\ap ')^\f32}{r^3} \spa 
r\equiv\s{\sum_{i=5}^9(x^i)^2},
\ea 
as follows 
\ba
&&ds^2=f^{-\f12}\Bigl(-(dt+\beta\sum_{i=1}^4 J_{ij}x^i dx^j)^2
+\sum_{i=1}^4{(dx^i)^2}\Bigr)+f^{\f12}\sum_{i=5}^9(dx^i)^2,\no
&&e^{2\phi}=f^{-\f12},\ 
\ \ \ F_{12}=F_{34}=-2\beta,\ \ \ \ H_{129}=H_{349}=-2\beta,\ \ \no
&& F_{0129}=F_{0349}=2\beta,\ \ \ \ 
F_{ijkl}=-\ep_{ijklm}\di_{m}f,\ (5\leq i,j,k,l,m \leq 9),
\ea
where $y=x^9$ and $J_{12}=J_{34}=1$.
We have checked all components of the EOMs and found that
this is indeed a supergravity solution.\footnote{To be exact we cannot
express the solution by using the field strengths $F^{(4)},H^{(3)},F^{(2)}$
in a gauge invariant way. This is because the action include the term 
$|\ti{F^{(4)}}|^2$ with $\ti{F}^{(4)}=F^{(4)}+A^{(1)}H^{(3)}$ and 
the gauge transformations of the potentials are given by 
$\delta A^{(1)}=d\lambda,\ \ \delta A^{(3)}=-\lambda H^{(3)}$. 
Thus we should specify the value of gauge field $A^{(1)}$ and in this paper
we have set $A^{(1)}=-\beta\sum_{i=1}^{4}J_{ij}x^idx^j$.} 

We can also consider the M-theory lift of this solution. 
As discussed before we know that the lift of the \godel Universe itself leads
to the pp-wave in M-theory which preserves 24 supersymmetries 
(\ref{metmpp24}) and (\ref{fluxmpp24}).
After changing the coordinates according to (\ref{genct}) 
we obtain the (localized) M5-brane solution in the pp-wave (the smeared
solution can be found in \cite{Singh:2002yd})
\ba
&&ds^2=f^{-\f13}\Bigl(-2dx^+dx^- -\beta^2\sum_{i=1}^4(\ti{x}^i)^2(dx^+)^2
+\sum_{i=1}^4(d\ti{x}^i)^2\Bigr)+f^{\f23}\sum_{i=5}^{9}(dx^i)^2,\no
&& F_{+129}=F_{+349}=2\beta,\ \ \ \ F_{ijkl}=-\ep_{ijklm}\di_{m}f
\ \ (5\leq i,j,k,l\leq 9).
\ea

It would be interesting to construct other
Dp-brane solutions in this \godel Universe or in other backgrounds (see 
\cite{Bain:2002nq} for
classical D-brane solutions in the maximally supersymmetric pp-wave).

\section{Discussion and Conclusions}


In this paper we have found several new supersymmetric 
\godel Universe backgrounds of string and M-theory. 
We discovered that not only T-duality, but also the
type-IIA/M-theory S-duality can give supersymmetric
\godel Universes from pp-waves.
We explained that the S-duality could be considered
equivalent to the T-duality when $n \leq 3$ by
considering M-theory on a three-torus. 
We found that there exist 
three inequivalent \godel Universes with 20
supersymmetries, one of which demonstrated the fact that the
S-duality transformation is not always equivalent to the
T-duality. 

We have found an intriguing M-theory $n=5$ \godel Universe with 
18 supersymmetries. This background does not seem to be related
to pp-waves by dualities. It would be interesting to consider if
one can compactify the background to a type IIA string theory 
solution, perhaps this can reveal connections to other 
supersymmetric backgrounds. 
In connection with the $n=5$ background with 18 supersymmetries
we found that it is a member of 
a family of M-theory \godel Universes with at least 16 supersymmetries
of which the $n=2$ and the $n=4$ M-theory \godel Universe
with 20 supersymmetries also are members.

We found two new supersymmetric backgrounds 
which are mixtures of \godel Universes and pp-waves. 
Perhaps one can use these backgrounds to construct a holographic
dual to the string theory since we can consider a small
deformation away from the pp-wave and try to deform the
gauge theory dual of the pp-wave \cite{Berenstein:2002jq} so that
it is dual to the deformed background.

We have also considered the string theory on two of the
type IIA \godel Universes with 20 supersymmetries.
We have shown that the string theory on these backgrounds
are solvable and found the string spectrum. 
This was done using the fact that one can light-cone
quantize the T-dual pp-wave when compactified in the
T-duality direction.
We note that part of the spectrum looks very much like
that found in 
\cite{Michelson:2002wa,Bertolini:2002nr} where a space-like
compactification of the maximally symmetric type IIB
pp-wave was considered. Perhaps one can find a similar Penrose
limit of $\ads_5 \times S^5$ as in \cite{Bertolini:2002nr} giving
directly the maximally supersymmetric type IIB pp-wave
\eqref{metmaxpp}-\eqref{fluxmaxpp} in the coordinate
system of \eqref{aftercoII}. This would enable one to find
a dual gauge theory description of the \godel Universe 
background.

We have also considered the spectrum of 
D-branes on the two backgrounds in which 
we quantized the string theory. 
Using again the T-dual pp-wave backgrounds we found the
spectrum of D-branes on one of the backgrounds,
and found a supergravity solution for a D4-brane 
on the other \godel Universe background.

In this paper we have not directly
addressed the perhaps most important physical question for
string theory on a background of the \godel Universe type,
namely whether string theory is consistent on these backgrounds
or not. Can string theory work when we have closed time-like curves?

We have seen several indications in this paper which suggest
that the supersymmetric \godel Universes
are consistent string theory backgrounds. The high amount of 
supersymmetry for some of these backgrounds alone suggests that
the backgrounds should be well-behaved. 
In particular they should not have closed string tachyons 
and should therefore be stable as string theory backgrounds.
Moreover, the fact that string theory on some of the \godel Universe
backgrounds is highly solvable, i.e. that
we can 1st-quantize the string theory and obtain
the string spectrum without problems, also suggests that the
string theory is well-defined.
Finally, the fascinating fact that there are 
type IIA \godel Universes which are S-dual to M-theory pp-waves
means physically that string theory on those type IIA \godel
backgrounds is well-behaved at strong coupling.


\ \newline
\noindent {\Large \bf Acknowledgments}
\vskip 0.1cm
\noindent We would like to thank N. Itzhaki, S. Minwalla, N. Obers,
H. Reall, N. Toumbas and A. Tseytlin
for useful discussions and correspondence.
TH thanks the Niels Bohr Institute for hospitality during part of this
work.
This work was supported by the DOE grant DE-FG02-91ER40654.

\begin{appendix}

\section{T and S-duality conventions}
\label{appconv}

\subsubsection*{T-duality conventions}

Consider a metric $g_{\mu \nu}$ which has a coordinate $y$
for which $g_{yy} = 1$. Consider moreover the NSNS 2-form
potential $B_{\mu \nu}$.
A T-duality in the $y$-direction is given by
\begin{equation}
\tilde{g}_{yy} = 1 \spa
\tilde{g}_{\mu \nu} = g_{\mu \nu} - g_{\mu y} g_{\nu y} 
+ B_{\mu y} B_{\nu y}
\spa
\tilde{g}_{\mu y} = B_{\mu y}
\end{equation}
\begin{equation}
\tilde{B}_{\mu \nu} = B_{\mu \nu}
- B_{\mu y} g_{\nu y} + g_{\mu y} B_{\nu y}
\spa
\tilde{B}_{\mu y} = g_{\mu y}
\end{equation}
Here $\mu,\nu \neq y$ and $\tilde{g}_{\rho \sigma}$ and
$\tilde{B}_{\rho \sigma}$ are the T-dual fields.
Note that a constant dilaton is mapped to a constant dilaton,
i.e. $\tilde{\phi} = \phi$.
If $y$ is on a circle of radius $R$ before the T-duality
then it is on a circle of radius $\alpha'/R$ after the T-duality.
For RR field strengths the T-duality transformations from 
IIA to IIB are
\begin{equation}
\tilde{F}_{\mu \nu \sigma \xi y} = F_{\mu \nu \sigma \xi}
\spa
\tilde{F}_{\mu \nu \sigma \xi \kappa} = F_{\mu \nu \sigma \xi \kappa y}
\spa
\tilde{F}_{\mu \nu y} = - F_{\mu \nu}
\spa
\tilde{F}_{\mu \nu \xi} = F_{\mu \nu \xi y}
\end{equation}
while for IIB to IIA they are
\begin{equation}
\tilde{F}_{\mu \nu \sigma \xi \kappa y} = F_{\mu \nu \sigma \xi \kappa}
\spa
\tilde{F}_{\mu \nu \sigma \xi} = F_{\mu \nu \sigma \xi y}
\spa
\tilde{F}_{\mu \nu \sigma y} = F_{\mu \nu \sigma}
\spa
\tilde{F}_{\mu \nu} = - F_{\mu \nu y}
\end{equation}
with $\mu, \nu, \sigma, \xi, \kappa \neq y$.
Note that we have the convention that
$F^{\mu_1 \cdots \mu_4} = \sqrt{-g} \epsilon^{\mu_1 \cdots \mu_{10}}
F_{\mu_5 \cdots \mu_{10}}$ with $\epsilon^{01 \cdots 9} = 1$.

\subsubsection*{M/IIA S-duality conventions}

Consider an M-theory background with the coordinate
$u$ being an explicit space-like isometry with $g_{uu}=1$.
We consider the S-duality between type IIA string theory
and M-theory with $u$ being the eleventh direction.
The relation between the eleven- and ten-dimensional metrics is
\begin{equation}
\label{metMIIA}
ds_{\rm M}^2 = ds_{\rm IIA}^2 + (du + A_\mu dx^\mu)^2
\end{equation}
where $A_\mu$ is the one-form RR gauge
potential in type IIA string theory. The dilaton is constant
since we assumed $g_{uu} = 1$. The relations between the
four-form field strength in M-theory and the RR four-form and NS-NS
three-form field strength in type IIA are
\begin{equation}
F^{\rm (M)}_{\mu \nu \sigma u} = H_{\mu \nu \sigma}
\spa
F^{\rm (M)}_{\mu \nu \sigma \xi}
= F^{\rm (IIA)}_{\mu \nu \sigma \xi}
\end{equation}
%

\section{Computations of supersymmetry}
\label{appsusy}

We compute the supersymmetry of various solutions found in
Section \ref{secsolutions}.

\subsection{Supersymmetry of compactified pp-waves}

We consider here the supersymmetry of
pp-waves of type IIB which have a compact direction.
We follow the discussion of
\cite{Blau:2001ne,Cvetic:2002hi,Michelson:2002wa,Bena:2002kq}.
In general the supersymmetry variation of the dilatino
$\lambda$ and the gravitino $\psi_\mu$ can be written%
\footnote{In this section we use the notation and conventions of
\cite{Bena:2002kq} so the definition of $\Gamma^\mu$ here is different
from the rest of the paper. For example
\[ \Gamma^3 \sigma_1 \vecto{\psi_1}{\psi_2}
= \Gamma^3 \vecto{\psi_2}{\psi_1}
= \vecto{\Gamma^3 \psi_2}{\Gamma^3 \psi_1} \] and so forth.
Here $\psi_1$,$\psi_2$ are two
Majorana spinors with same chirality both with 16 components.
We also use that $\Gamma^+ = \Gamma^0 + \Gamma^9$,
$\Gamma^- = \frac{1}{2} ( \Gamma^0 - \Gamma^9 )$ and
$\Gamma^{12345678} = \Gamma^{09}$.}
\begin{equation}
\label{susyvar}
\delta \lambda = \Gamma^+ W \eta
\spa
\delta \psi_\mu = D_\mu \eta + \Omega_\mu \eta
\end{equation}
where $\Omega_\mu$ is the torsion in type IIB. $W$ and $\Omega_\mu$ involves
both NS-NS and RR field strengths. We use the conventions
of \cite{Bena:2002kq} for the supersymmetry variations \eqref{susyvar}
in type IIB.
We then compactify the solutions along the Killing vector with unit norm
\begin{equation}
\label{killvec}
\xi = \partial_+ - \frac{1}{2} \partial_- + \beta^2 \sum_{i<j} J_{ij} M_{ij}
\end{equation}
where $M_{ij} = \tilde{x}^i \partial_j - \tilde{x}^j \partial_i$.
This leads to the extra condition on Killing spinors
\begin{equation}
Q \eta = 0 \spa Q \equiv \Omega_+
+ \frac{1}{2} \sum_{i<j} J_{ij} \Gamma^{ij}
\end{equation}
where we have put $\beta=1$ for simplicity.
To further simplify the analysis we define
\begin{equation}
J_1 = i \sigma_2 \Gamma^{12} \spa
J_2 = i \sigma_2 \Gamma^{34} \spa
J_3 = i \sigma_2 \Gamma^{56} \spa
J_4 = i \sigma_2 \Gamma^{78}
\end{equation}
We see that the eigenvalues of $J_i$ are $\pm 1$.
Note that $J_1 J_2 J_3 J_4$ has eigenvalue $1$ for kinematical
supersymmetries, i.e. when $ \Gamma^+ \eta = 0$, and eigenvalue
$-1$ for dynamical supersymmetries, i.e. when $ \Gamma^- \eta = 0$.
Note also that any eigenvalue of $(J_1,J_2,J_3,J_4)$ has degeneracy two.

We consider now the maximally supersymmetric pp-wave of type IIB
\eqref{metmaxpp}-\eqref{fluxmaxpp} \cite{Blau:2001ne}
with 32 supersymmetries.
We have
\begin{eqnarray}
\Omega_+ &=& - \frac{1}{4} ( \Gamma^{1234} + \Gamma^{5678})
\Gamma^+ \Gamma^- (i \sigma_2)
= i \sigma_2 \frac{1}{4} ( J_1 J_2 + J_3 J_4 ) ( 1 + J_1 J_2 J_3 J_4 )
\nn \\
Q &=&i \sigma_2 \left[ \frac{1}{4}
( J_1 J_2 + J_3 J_4 ) ( 1 + J_1 J_2 J_3 J_4
)
- \frac{1}{2} ( J_1 + J_2 + J_3 - J_4 ) \right]
\end{eqnarray}
When the direction \eqref{killvec} is compact, we see that we have
8 kinematical and 12 dynamical supersymmetries, giving altogether
20 supersymmetries.

For the mixed pp-wave solution with NS-NS and RR three-form flux
\eqref{metppintD3}, \eqref{fluxppmix} \cite{Berenstein:2002jq} we
compute
\begin{eqnarray}
W &=& - \frac{1}{2} \Big[ \sigma_1 \sin \gamma + \sigma_3 \cos \gamma \Big]
( J_1 + J_2 ) \nn \\
\Omega_+ &=& - \frac{1}{4} \Big[ 2\sigma_1 \sin \gamma -
(1 + J_1 J_2 J_3 J_4 ) \sigma_3 \cos \gamma \Big] ( J_1 + J_2 )
\nn \\
Q & = & - \frac{1}{4} \Big[ 2\sigma_1 \sin \gamma -
(1 + J_1 J_2 J_3 J_4 ) \sigma_3 \cos \gamma - 2 i \sigma_2 \Big]
( J_1 + J_2 )
\end{eqnarray}
We see that the dilatino variation $\Gamma^+ W \eta = 0$
gives 24 supersymmetries in the uncompactified case, i.e. 16 kinematical
and 8 dynamical.
When the direction \eqref{killvec} is compact we see that
we need $J_1 + J_2 = 0$ for a generic value of $\gamma$, thus giving
16 supersymmetries, i.e. 8 kinematical and 8 dynamical.
If $\gamma = 0$ then we get instead 12 kinematical and 8 dynamical
supersymmetries, giving altogether 20 supersymmetries.

For the mixed pp-wave solution with RR three and five-form flux
\eqref{metpp28}-\eqref{fluxpp28}
\cite{Bena:2002kq} we have
\begin{eqnarray}
W &=& - \frac{1}{2} \sin \gamma \,
\sigma_3 ( J_1 + J_2 + J_3 - J_4 ) \nn \\
\Omega_+ &=& \frac{1}{4} i \sigma_2
\Big[ (J_1 J_2 + J_3 J_4 ) \cos \gamma
- (J_1 + J_2 + J_3 - J_4 ) \sigma_1 \sin \gamma \Big] ( 1 + J_1 J_2 J_3 J_4)
\nn \\
Q &=&i \sigma_2 \left[ \frac{1}{4}
\Big( (J_1 J_2 + J_3 J_4 ) \cos \gamma
- (J_1 + J_2 + J_3 - J_4 ) \sigma_1 \sin \gamma \Big)
( 1 + J_1 J_2 J_3 J_4) \right. \nn \\
&& \left. - \frac{1}{2} ( J_1 + J_2 + J_3 - J_4 ) \right] \label{mixedss}
\end{eqnarray}
We see that the dilatino variation $\Gamma^+ W \eta = 0$
gives 28 supersymmetries in the uncompactified case, i.e. 16 kinematical
and 12 dynamical.
When the direction \eqref{killvec} is compact we see that
we have 0 kinematical and 12 dynamical supersymmetries, giving altogether
12 supersymmetries. It is also easy to see that the special case
$\gamma=0,\pi/2$ there are 20 supersymmetries
(12 dynamical ones + 8 kinematical ones).

\subsection{Supersymmetry of M-theory backgrounds}

Here we show the analysis of spinors of several \godel backgrounds
in M-theory explicitly.

\noindent {\bf (1) M-theory \godel 
background} \eqref{metn4M1} and \eqref{fluxn4M1} (20 susy)

After we solve the Killing spinor 
equations (\ref{susycount}), we obtain the following 
ten
 constant spinors 
\ba
\eta^{(1)}=\eta^{--++},
\ \ \ \eta^{(2)}=\eta^{-++-},\ \ \ \eta^{(3)}=\eta^{+-+-}\ \ \ 
\eta^{(4)}=\eta^{--+-},\ \ \   \eta^{(5)}=\eta^{----},
\ea
where we specify by using the $\pm 1$ values of 
$(\gamma^{012},\gamma^{034},\gamma^{056},\gamma^{078})$ (note that
there is the degeneracy of factor two) as well as the other ten spinors
which depend on the coordinates $x^1,...,x^8$ linearly 
\ba
\eta^{(6)}&=&\left(1+2\beta J_{ij}\Gamma^{0i}x^j\right)\eta^{++-+},\no
\eta^{(7)}&=&\left(1-2\beta \Gamma^{04}x^3+2\beta\Gamma^{03}x^4
-2\beta\Gamma^{06}x^5+2\beta\Gamma^{05}x^6\right)\eta^{+-++},\no
\eta^{(8)}&=&\left(1-2\beta\Gamma^{02}x^1+2\beta\Gamma^{01}x^2
-2\beta\Gamma^{06}x^5+2\beta\Gamma^{05}x^6\right)\eta^{-+++},\no
\eta^{(9)}&=&\left(1-2\beta\Gamma^{04}x^3+2\beta\Gamma^{03}x^4
+2\beta\Gamma^{08}x^7-2\beta\Gamma^{07}x^8\right)\eta^{+---},\no
\eta^{(10)}&=&\left(1-2\beta\Gamma^{02}x^1+2\beta\Gamma^{01}x^2
+2\beta\Gamma^{08}x^7+2\beta\Gamma^{07}x^8\right)\eta^{-+--}.\label{ks420}
\ea
Thus we can conclude that there are 20 supersymmetries in this background.\\

\noindent {\bf (2) M-theory \godel 
background} \eqref{metn5},\eqref{fluxn5} and \eqref{almost} (18 or 16 susy)

In this case again we have only to solve (\ref{susycount}). We find
the constant 10 Killing spinors 
\ba
\chi^{(1)}=\chi^{+-++},\ \ \chi^{(2)}=\chi^{-+++}, \ \ 
\chi^{(3)}=\chi^{--++},\ \ \chi^{(4)}=\chi^{--+-},\ \ 
\chi^{(5)}=\chi^{---+},
\ea
and the non-constant 6 Killing spinors
\ba
\chi^{(6)}&=&\left(1+2\beta J_{ij}\Gamma^{0i}x^j\right)\chi^{++--},\no
\chi^{(7)}&=&\left(1-2k\beta \Gamma^{04}x^3+2k\beta\Gamma^{03}x^4
-2k\beta\Gamma^{06}x^5+2k\beta\Gamma^{05}x^6\right)\chi^{+--+},\no
\chi^{(8)}&=&\left(1-2k\beta\Gamma^{04}x^3+2k\beta\Gamma^{03}x^4
-2k\beta\Gamma^{08}x^7+2k\beta\Gamma^{07}x^8\right)\chi^{+-+-},\label{k678}
\ea
where the values of $(J_{12},J_{34},J_{56},J_{78},J_{9\sharp})$ is given by
$(k+1,k,k,k,1)$ as defined in (\ref{almost}).
Thus we can conclude that there are 16 supersymmetries in this model.

If we assume $k=1$, we get the extra Killing spinor
\ba
\chi^{(9)}&=&\left(1-2\beta\Gamma^{04}x^3+2\beta\Gamma^{03}x^4
+2\beta\Gamma^{0\sharp}x^9-2\beta\Gamma^{09}x^{\sharp}\right)\chi^{+---}.
\label{k9}
\ea
Therefore in this special case $k=1$ we get 18 supersymmetries.\\

\noindent {\bf (3) M-theory \godel
background} \eqref{metn4M2}-\eqref{fluxn4M2} (12 susy) 

Since we have already shown that the small radius limit (type IIA model)
has 12 supersymmetries (see (\ref{mixedss})) we have only to check that
the background in M-theory does not allow more than 12 supersymmetries.
To see this let us consider the time component $\mu=0$
of Killing spinor equation (\ref{killspin}). This is given by
\ba
\label{fdf}
&&\left(\di_0 +\f{\beta}{2}(\Gamma^{12}
+\Gamma^{34}+\Gamma^{56}-\Gamma^{78})\right)\eta  
= \f12 \beta\cos\gamma(\Gamma^{01234}+\Gamma^{05678})\eta
\\ &&
-\f16 \beta(\Gamma^{0129\sharp}+\Gamma^{0349\sharp}
+\Gamma^{0569\sharp}-\Gamma^{0789\sharp})\eta
+\f13\beta\sin\gamma(\Gamma^{129}
+\Gamma^{349}+\Gamma^{569}-\Gamma^{789})\eta. \nn
\ea
For generic $\gamma$ ($\gamma\neq 0,\pi/2$), we have the conditions
$(\Gamma^{1234}+\Gamma^{5678})\eta=(\Gamma^{012}
+\Gamma^{034}+\Gamma^{056}-\Gamma^{078})\eta=0$. Thus we
find 12 supersymmetries in this background.\\

\noindent {\bf (4) M-theory \godel
background} \eqref{metn2mix}-\eqref{fluxn2mix} (16 susy)

The Killing spinor can be examined as in the previous case (\ref{fdf}).
The result is given by the constraint $(\Gamma^{12}+\Gamma^{34})\eta=0$,
and thus we have 16 supersymmetries generically.

\end{appendix}

\addcontentsline{toc}{section}{References}




\providecommand{\href}[2]{#2}\begingroup\raggedright\endgroup

\end{document}